\documentclass[12pt, preprint]{aastex}

\newcommand{\mum}{$\rm \, \mu m$}
\newcommand{\HII}{\ion{H}{2}~}
\def\pht{\rm ISO--PHT--S~}

\shorttitle{PAHs as a tracer of star formation?}
\shortauthors{Peeters, E, Spoon, H.W.W. and Tielens, A.G.G.M.}

\begin{document}

\title{PAHs as a tracer of star formation?}

\author{E. Peeters}
\affil{NASA-Ames Research Center, Mail Stop 245-6, Moffett
Field, CA 94035, USA}
\email{epeeters@mail.arc.nasa.gov}
\author{H.W.W. Spoon}
\affil{Kapteyn Institute, P.O. Box 800, 9700 AV Groningen, The Netherlands;\\ Cornell University, Astronomy Department, Ithaca, NY 14853, USA}
\and
\author{A.G.G.M. Tielens}
\affil{SRON National Institute for Space Research/Kapteyn
Institute, P.O. Box 800, 9700 AV Groningen, The Netherlands}

\begin{abstract}
Infrared (IR) emission features at 3.3, 6.2, 7.7, 8.6 and 11.3\,$\mu$m
are generally attributed to IR fluorescence from (mainly) FUV pumped
large Polycyclic Aromatic Hydrocarbon (PAH) molecules. As such, these
features trace the FUV stellar flux and are thus a measure of star
formation. We examined the IR spectral characteristics of Galactic
massive star forming regions and of normal and starburst galaxies, as
well as AGNs and ULIRGs. The goal of this study is to analyze if PAH
features are a good qualitative and/or quantitative tracer of star
formation and hence the application of PAH emission as a diagnostic
tool in order to identify the dominant processes contributing to the
infrared emission from Seyfert's and ULIRGs.  We develop a new MIR/FIR
diagnostic diagram based upon our Galactic sample and compare it to
the diagnostic tools of \citet{Genzel98} and \citep{Laurent00}, with
these diagnostic tools also applied to our Galactic sample.  This
MIR/FIR diagnostic is derived from the FIR normalized 6.2\,$\mu$m PAH
flux and the FIR normalized 6.2\,$\mu$m continuum flux. Within this
diagram, the Galactic sources form a sequence spanning a range of 3
orders of magnitude in these ratios, ranging from embedded compact
\HII regions to exposed Photo Dissociation Regions (PDRs) and the
(diffuse) ISM.  However, the variation in the 6.2\,$\mu$m PAH
feature-to-continuum ratio is relative small. Comparison of our
extragalactic sample with our Galactic sources revealed an excellent
resemblance of normal and starburst galaxies to exposed PDRs.  While
Seyfert-2's coincide with the starburst trend, Seyfert-1's are
displaced by at least a factor 10 in 6.2\,$\mu$m continuum flux, in
accordance with general orientation dependent unification schemes for
AGNs.  ULIRGs show a diverse spectral appearance. Some show a typical
AGN hot dust continuum. More, however, are either starburst-like or
show signs of strong dust obscuration in the nucleus.  One
characteristic of the ULIRGs also seems to be the presence of more
prominent FIR emission than either starburst galaxies or AGNs.  We
discuss the observed variation in the Galactic sample in view of the
evolutionary state and the PAH/dust abundance and discuss the use of
PAHs as quantitative tracers of star formation activity. Based on
these investigations we find that PAHs may be better suited as a
tracer of B stars, which dominate the Galactic stellar energy budget,
than as a tracer of massive star formation (O stars).
\end{abstract}


\section{Introduction}

The mid-infrared (MIR) spectra of many objects with associated dust
and gas are dominated by the well-known emission features at 3.3, 6.2,
7.7, 8.6 and 11.2~$\mu$m commonly called the unidentified infrared
(UIR) bands.  These bands are now generally attributed to vibrational
emission of Polycyclic Aromatic Hydrocarbons (PAHs) containing
$\simeq$ 50 carbon atoms \citep{Leger:84,
Allamandola:autoexhaust:85,Allamandola:rev:89, Puget:revpah:89,
Tielens:parijs:99, Tielens:korea:00}. One key aspect of these IR
emission features is that they are particularly bright in regions
illuminated by UV bright, early type stars responsible for \HII
regions and reflection nebulae (RNe).

With the launch of the Infrared Space Observatory (ISO), a huge amount
of IR data became available, revealing the omnipresent
nature of these PAH features.  By now, these features have been
detected in a wide range of objects and environments, from post-AGB
stars and planetary nebulae (PNe), to \HII regions, RNe, the diffuse
interstellar medium (ISM) and extragalactic sources, up to redshifts
of $z$=0.3. Various studies of Galactic sources with bright PAH
emission features have been performed (see e.g. the ISO special issue,
A\&A, 315, 1996 and, for a recent review, \cite{Peeters:review:04}).
Of particular importance here are the studies of the PAHs in 
\HII regions, which characterized the UIR bands (and their
variations) in massive star forming regions \citep{Verstraete:m17:96,
VanKerckhoven:plat:00, Hony:oops:01, Peeters02b, Vermeij:pahs:01,
Verstraete01, vanDiedenhoven:chvscc:03}.

The MIR extragalactic ISO studies established that PAH emission
features in extragalactic environments are very similar to those in
Galactic star forming regions \citep[e.g.][]{Genzel98, Lutz98,
Mirabel98, Charmandaris99, Rigopoulou99,Clavel00, Helou00, Tran01}.
This property has since been used qualitatively and quantitatively as
diagnostics for the ultimate physical processes powering Galactic
nuclei.
In particular, \cite{Genzel98} found that the 7.7 \mum\, PAH
feature-to-continuum ratio is on average an order of magnitude smaller
for Seyfert 1 galaxies (Sf1's) than for starburst
galaxies. Conversely, the ratio of the high- to low-excitation
emission lines [\ion{O}{4}]/[\ion{Ne}{2}] was found to be two orders
of magnitude higher for Active Galactic Nuclei (AGNs, both Sf1's
and 2's) than for starburst galaxies. These two ratios are therefore
used to separate AGNs and starburst galaxies. Ultra-Luminous InfraRed
Galaxies (ULIRGs) are found to reside in between the two groups,
although closer to the starburst galaxies, indicating that 80\% of the
ULIRGs are predominantly powered by star formation.
A second MIR diagnostic plot was devised by \cite{Laurent00}. This
diagram separates \HII regions, PDRs and AGN-dominated
spectra on the basis of their distinctly different ratio of warm
(15\,$\mu$m) to hot (6\,$\mu$m) continuum and the value of their
6.2\,$\mu$m PAH feature-to-continuum ratio. The diagram thereby allows an 
estimation of the contribution of the AGN, PDRs and \HII regions to a given
MIR spectrum.
Given the absence of MIR emission lines for the majority
of ULIRGs, Lutz et al. (1998) adapted the Genzel
diagram and replaced the [OIV]/[NeII] line ratio by
the ratio of 5.9 to 60 \mum\, continuum
(5.9/60 ratio). In this diagram the majority of ULIRGs are found at
high PAH feature-to-continuum ratio and low 5.9/60 ratio, close to the
position of starburst galaxies.  Fewer ULIRGs are found along the
mixing line which extends to pure AGNs, at low PAH
feature-to-continuum ratio and high 5.9/60 ratio.
\cite{Clavel00} compared the MIR spectral properties of Sf1s and
Sf2s. They found that the 7.7 \mum\, PAH luminosity
distributions of both types are the same, indicating that the
properties of the host galaxies are unrelated to the type-1/2
classification of the nuclear activity. They also found that, on
average, the 7.7\,$\mu$m continuum of Sf2s is a factor $\sim$8 weaker
than of Sf1s and attribute this to dust obscuration in the nuclear
environment, which, in line with the unified models for AGNs
\citep{Antonucci93}, is due to the orientation of the AGN torus.
Another result of this study is that except for the emission line
spectrum, the 3--12\,$\mu$m spectra of Sf2s appear very
similar to those of normal and starburst galaxies.

The goal of this paper is to establish the characteristics of PAH
emission bands in regions of massive star formation in the Milky Way,
in order to use them as a tool for studying star formation on a
galaxy-wide scale and to apply these tools to the study of extragalactic
starbursts in Seyferts, ULIRGs and QSOs.

Sect.\,\ref{sample} presents MIR observations of a Galactic sample and
of a sample of normal, starburst, Seyfert and (ultra-luminous) IR
galaxies and QSOs.  Their spectral characteristics are discussed in
Sect.\,\ref{spchar}.  In Sect.\,\ref{diagnostictools}, we investigate
three IR diagnostic tools designed to distinguish AGN- from
starburst-dominated spectra. Sect.\,\ref{discussion} highlights the
PAH abundance as fraction of the total amount of dust and the
application of the PAH emission bands as tracers of star formation.
The conclusions are stated in Sect.\,\ref{conclusions}.

\section{Observations}
\label{sample}

\subsection{Galactic sample}
\label{sample_galactic}

The sample of \HII regions was taken from the Galactic ``Ultra Compact
\HII region'' ISO program \citep{Peeters:cataloog:01} complemented with the
Orion Bar, M17 and 30\,Dor.  Only those spectra with sufficient S/N
are included in this sample. As a reference, we included RNe, heavily
embedded protostars exhibiting PAH emission and the
(diffuse) ISM in various regions. For details on the sources and
references, see Table \ref{log}.

Most spectra were obtained with the Short Wavelength Spectrometer
\citep[SWS,][]{deGraauw96} on board ISO \citep{Kessler96} using the
AOT~01 scanning mode at various speeds with resolving power
($\lambda/\Delta \lambda$) ranging from 400 to 1500 (see Table
\ref{log}). The data were processed with the SWS Interactive Analysis
package IA$^3$ \citep{deGraauw96} using calibration files and
procedures equivalent with pipeline version 10.0 or later.  A detailed
account of this reduction can be found in \citet{Peeters:cataloog:01}. We
spliced the subbands to form a continuous spectrum from 5 to 15
\mum. The amount of shifting between subbands is within the calibration
uncertainties for the region of interest and does influence
the derived fluxes.  A few sources were obtained with \pht,
ISO--CAM--CVF, IRTS--MIRS and MSX with resolving power
($\lambda/\Delta \lambda$) of respectively $\sim$ 90, 35, 30 and
800. We refer to the original papers (see Table \ref{log}) for an
account of the reduction process.

\clearpage


\begin{deluxetable}{llll@{\hspace{9pt}}c@{\hspace{9pt}}l@{\hspace{14pt}}ccccccl}
\tabletypesize{\scriptsize}
\rotate
\tablewidth{0pt}
\tablecolumns{12}
\tablecaption{Journal of observations together with the derived fluxes.\label{log} }
\tablehead{
\colhead{Source} & \colhead{$\alpha$(J2000)\tablenotemark{a}} & \colhead{$\delta$(J2000)\tablenotemark{a}} & 
\colhead{Instrument\tablenotemark{b}} & \colhead{TDT\tablenotemark{c}} & \colhead{Ref.} & 
\colhead{6.2 PAH\tablenotemark{d}} & \colhead{cont.\tablenotemark{d}} & \colhead{cont.\tablenotemark{d}} & 
\colhead{cont.\tablenotemark{d}} & \colhead{FIR\tablenotemark{d}} & \colhead{Scaling\tablenotemark{e}} & \colhead{Ref.} \\ 
\colhead{} & \colhead{} & \colhead{} & \colhead{} & \colhead{} & \colhead{} & \colhead{} & \colhead{[5.3,5.8]} & \colhead{[6.0,6.5]} 
& \colhead{[14,15]} & \colhead{} & & \colhead{FIR}
}
\startdata

\multicolumn{13}{c}{{\footnotesize HII regions}}\\
 &  &  &  &  &  & W/m$^2$ & W/m$^2$ & W/m$^2$ & W/m$^2$ & W/m$^2$ & & \\[10pt]

W~3A 02219+6125\tablenotemark{f}
                           & 02 25 44.59 & $+$62 06 11.20 & ISO-SWS~01(2) & 64600609 &  1  & 2.16(-13) & 3.77(-13) & 4.99(-13) & 4.04(-12) & 1.56(-09) & 0.33 & 1 \\*

                           &             &                & ISO-SWS~01(2) & 78800709 &  1  & 2.36(-13) & 3.38(-13) & 4.87(-13) & 4.38(-12) & 1.56(-09) & 0.33 & 1 \\ 

30~Dor                     & 05 38 46.00 & $-$69 05 07.91 & ISO-SWS~01(4) & 17100512 &  2  & 2.61(-14) & 7.18(-14) & 4.84(-14) & 3.08(-13) & 5.75(-11) & 0.29 & 18\\

OrionBar~D8                & 05 35 18.22 & $-$05 24 39.89 & ISO-SWS~01(2) & 69501409 &  3  & 3.16(-13) & 2.05(-13) & 3.30(-13) & 1.72(-12) & -& &   \\*

OrionBar~BRGA              & 05 35 19.31 & $-$05 24 59.90 & ISO-SWS~01(2) & 69502108 &  -  & 3.76(-13) & 2.10(-13) & 3.31(-13) & 1.07(-12) & -& &    \\*

OrionBar~D5                & 05 35 19.81 & $-$05 25 09.98 & ISO-SWS~01(2) & 83101507 &  -  & 5.44(-13) & 2.34(-13) & 3.82(-13) & 1.01(-12) & 3.50(-11) &  -    &19   \\*

OrionBar~H2S1              & 05 35 20.31 & $-$05 25 19.99 & ISO-SWS~01(2) & 69501806 &  4  & 3.47(-13) & 1.89(-13) & 3.04(-13) & 6.94(-13) & -& &    \\*

OrionBar~D2                & 05 35 21.40 & $-$05 25 40.12 & ISO-SWS~01(2) & 69502005 &  -  & 1.20(-13) & 9.23(-14) & 1.20(-13) & 4.67(-13) & -& &    \\*

Orion                      & \multicolumn{2}{l}{centered on Trapezium} & MSX & -     &  5  & 1.77(-5)\tablenotemark{g} & 6.50(-06)\tablenotemark{g} & 1.51(-5)\tablenotemark{g} & 5.52(-05)\tablenotemark{g} & 2.80(-3)\tablenotemark{g,h} & - & 20 \\

IRAS~10589$-$6034          & 11 00 59.78 & $-$60 50 27.10 & ISO-SWS~01(2) & 26800760 &  1  & 1.28(-13) & 1.05(-13) & 1.38(-13) & 4.46(-13) & 1.47(-10) & 0.74 & 1 \\

IRAS~12063$-$6259          & 12 09 01.15 & $-$63 15 54.68 & ISO-SWS~01(2) & 25901414 &  1  & 1.18(-13) & 1.57(-13) & 1.96(-13) & 5.79(-13) & 1.60(-10) & 0.81 & 1 \\

IRAS~12073$-$6233\tablenotemark{f}
                           & 12 10 00.32 & $-$62 49 56.50 & ISO-SWS~01(2) & 25901572 &  1  & 7.76(-14) & 5.14(-13) & 6.12(-13) & 7.75(-12) & 1.10(-09) & 0.44 & 1 \\

IRAS~12331$-$6134\tablenotemark{f} 
                           & 12 36 01.9  & $-$61 51 03.9  & ISO-SWS~01(2) & 29900470 &  1  & 3.57(-14) & 4.32(-14) & 5.47(-14) & 1.52(-13) & 1.34(-10) & 0.18 & 1 \\

IRAS~15384$-$5348\tablenotemark{f} 
                           & 15 42 17.16 & $-$53 58 31.51 & ISO-SWS~01(2) & 29900661 &  1  & 3.26(-13) & 1.53(-13) & 2.55(-13) & 6.50(-13) & 4.43(-10) & 0.50 & 1 \\

IRAS~15502$-$5302          & 15 54 05.99 & $-$53 11 36.38 & ISO-SWS~01(2) & 27301117 &  1  & 9.90(-14) & 2.32(-13) & 2.88(-13) & 9.73(-13) & 8.53(-10) & 0.63 & 1 \\

IRAS~16128$-$5109\tablenotemark{f} 
                           & 16 16 39.3  & $-$51 16 58.3  & ISO-SWS~01(2) & 29402233 &  1  & 8.48(-14) & 7.20(-14) & 8.13(-14) & 2.23(-13) & 5.33(-10) & 0.21 & 1 \\

IRAS~17160$-$3707\tablenotemark{f} 
                           & 17 19 26.1  & $-$37 10 53.8  & ISO-SWS~01(2) & 32400821 &  1  & 4.37(-14) & 5.54(-14) & 6.82(-14) & 1.45(-13) & 3.51(-10) & 0.25 & 1 \\

IRAS~17221$-$3619\tablenotemark{f} 
                           & 17 25 31.7  & $-$36 21 53.5  & ISO-SWS~01(2) & 33100380 &  1  & 7.46(-14) & 5.36(-14) & 7.24(-14) & 2.38(-13) & 2.08(-10) & 0.23 & 1 \\

IRAS~17279$-$3350\tablenotemark{f} 
                           & 17 31 18.0  & $-$33 52 49.4  & ISO-SWS~01(2) & 32200877 &  1  & 9.74(-14) & 5.88(-14) & 8.29(-14) & 2.31(-13) & 1.28(-10) & 0.35 & 1 \\

Sgr~C\tablenotemark{f}             
                           & 17 44 35.6  & $-$29 27 29.3  & ISO-SWS~01(2) & 84100301 &  1  & 6.79(-14) & 3.82(-14) & 2.72(-14) & 2.19(-13) & -         &  -&   \\

IRAS~17455$-$2800          & 17 48 41.5  & $-$28 01 38.3  & ISO-SWS~01(2) & 28701327 &  1  & 1.15(-13) & 1.24(-13) & 1.81(-13) & 1.03(-12) & 2.29(-10) & 0.53 & 1 \\

IRAS~17591$-$2228          & 18 02 13.2  & $-$22 27 58.9  & ISO-SWS~01(2) & 51500580 &  1  & 2.70(-14) & 4.89(-14) & 4.19(-14) & 9.10(-14) & 7.26(-11) & 0.57 & 1 \\

IRAS~18032$-$2032          & 18 06 13.93 & $-$20 31 43.28 & ISO-SWS~01(2) & 51500478 &  1  & 1.09(-13) & 7.88(-14) & 1.03(-13) & 3.53(-13) & 3.19(-10) & 0.54 & 1 \\

IRAS~18116$-$1646          & 18 14 35.29 & $-$16 45 20.99 & ISO-SWS~01(2) & 70300302 &  1  & 1.62(-13) & 4.58(-14) & 1.01(-13) & 4.23(-13) & 2.94(-10) & 0.39 & 1 \\

GGD~-27 ILL\tablenotemark{i}        
                           & 18 19 12.00 & $-$20 47 31.10 & ISO-SWS~01(2) & 14802136 & 1,6 & 1.37(-13) & 2.81(-13) & 2.88(-13) & 5.21(-13) & 2.16(-10) & 1.   & 1 \\

M17 iram pos. 1            & 18 20 28.98 & $-$16 11 50.78 & ISO-SWS~01(2) & 10201811 &  7  & 7.54(-14) & 9.74(-14) & 1.57(-13) & 1.99(-12) & 4.82(-11) & -    & 21  \\
											
M17 iram pos. 2            & 18 20 27.59 & $-$16 12  0.90 & ISO-SWS~01(2) & 09900212 &  7  & 4.48(-14) & 1.44(-13) & 3.05(-13) & 4.07(-12) & -          & -    &  \\
											
M17 iram pos. 3            & 18 20 26.19 & $-$16 12 11.02 & ISO-SWS~01(2) & 09901413 &  7  & 1.20(-13) & 1.59(-13) & 2.18(-13) & 3.21(-12) & -          & -    &  \\
											
M17 iram pos. 4            & 18 20 24.79 & $-$16 12 21.10 & ISO-SWS~01(2) & 09900214 &  7  & 3.35(-13) & 1.59(-13) & 2.49(-13) & 1.28(-12) & -          & -    &  \\
											
M17 iram pos. 5            & 18 20 23.40 & $-$16 12 31.21 & ISO-SWS~01(2) & 09901415 &  7  & 2.68(-13) & 1.46(-13) & 2.02(-13) & 4.54(-13) & -          & -    &  \\
											
M17 iram pos. 6            & 18 20 22.09 & $-$16 12 41.29 & ISO-SWS~01(2) & 09900216 &  7  & 2.03(-13) & 8.34(-14) & 1.51(-13) & 2.26(-13) & -          & -    &  \\
											
                           &             &                &               & 32900866 &  7  & 2.05(-13) & 1.03(-13) & 1.52(-13) & 1.81(-13) & -          & -    &  \\
											
M17 iram pos. 7            & 18 20 20.70 & $-$16 12 51.41 & ISO-SWS~01(2) & 09901417 &  7  & 5.81(-14) & 2.11(-14) & 2.98(-14) & 8.30(-14) &  -         & -    &\\
											
M17 iram pos. 8            & 18 20 19.31 & $-$16 13 01.49 & ISO-SWS~01(2) & 09900218 &  7  & 5.43(-14) & 5.28(-14) & 6.69(-14) & 1.02(-13) & 1.22(-11) & -    &21  \\

M17 North                  & 18 20 32.77 & $-$16 01 42.49 & ISO-SWS~01(2) & 09901105 & 8   & 7.01(-14) & 1.03(-13) & 1.38(-13) & 1.03(-13) & 1.45(-10) &  0.23 &  \\

IRAS~18317$-$0757          & 18 34 24.94 & $-$07 54 47.92 & ISO-SWS~01(2) & 47801040 &  1  & 2.93(-13) & 1.51(-13) & 2.55(-13) & 7.28(-13) & 1.97(-10) &  0.78 & 1  \\

IRAS~18434$-$0242          & 18 46 04.09 & $-$02 39 20.02 & ISO-SWS~01(2) & 15201383 & 1,6 & 1.88(-13) & 3.58(-13) & 5.11(-13) & 4.41(-12) & 6.98(-10) &  0.92 & 1  \\

IRAS~18469$-$0132          & 18 49 33.0  & $-$01 29 03.70 & ISO-SWS~01(2) & 71100888 &  1  & 7.63(-14) & 6.59(-14) & 6.78(-14) & 2.11(-13) & 1.26(-10) &  0.66 & 1  \\

IRAS~18479$-$0005          & 18 50 30.8  & $-$00 01 59.40 & ISO-SWS~01(2) & 15201791 &  1  & 3.05(-14) & 9.97(-14) & 9.47(-14) & 5.81(-13) & 2.79(-10) &  0.77 & 1  \\

IRAS~18502$+$0051          & 18 52 50.21 & $+$00 55 27.59 & ISO-SWS~01(2) & 15201645 &  1  & 1.08(-13) & 9.13(-14) & 8.66(-14) & 5.91(-13) & 1.25(-10) &  0.86 & 1  \\

IRAS~19207$+$1410\tablenotemark{f}    
                           & 19 23 02.4  & $+$14 16 40.60 & ISO-SWS~01(2) & 15001041 &  1  & 6.05(-14) & 6.22(-14) & 5.12(-14) & 2.95(-13) & 3.79(-10) &  0.28 & 1  \\

IRAS~19442$+$2427\tablenotemark{i ?} 
                           & 19 46 20.09 & $+$24 35 29.40 & ISO-SWS~01(2) & 15000444 & 1,6 & 2.54(-13) & 2.46(-13) & 2.98(-13) & 6.24(-13) & 2.71(-10) &  0.68 & 1  \\

IRAS~19598$+$3324\tablenotemark{i}     
                           & 20 01 45.6  & $+$33 32 43.70 & ISO-SWS~01(4) & 38402466 &  1  & 1.41(-13) & 2.22(-12) & 2.32(-12) & 7.53(-12) & 8.89(-10) &  0.90 & 1  \\

IRAS~21190$+$5140          & 21 20 44.89 & $+$51 53 26.99 & ISO-SWS~01(2) & 15901853 &   1 & 9.38(-14) & 9.06(-14) & 1.10(-13) & 8.92(-13) & 8.66(-11) & 1.   & 1  \\

IRAS~22308$+$5812          & 22 32 45.95 & $+$58 28 21.00 & ISO-SWS~01(2) & 17701258 &  1,6& 1.14(-13) & 8.18(-14) & 1.12(-13) & 1.63(-13) & 7.39(-11) & 0.72 & 1  \\*

		           &             &                &               & 56101082 &  1  & 1.31(-13) & 8.03(-14) & 1.17(-13) & 1.22(-13) & 7.39(-11) & 0.72 & 1 \\ 

IRAS~23030$+$5958          & 23 05 10.60 & $+$60 14 40.99 & ISO-SWS~01(2) & 22000961 &  1  & 7.79(-14) & 7.59(-14) & 8.54(-14) & 2.09(-13) & 1.25(-10) & 0.57 & 1  \\

IRAS~23133$+$6050          & 23 15 31.39 & $+$61 07 08.00 & ISO-SWS~01(2) & 22001506 &  1  & 2.45(-13) & 1.37(-13) & 2.13(-13) & 4.92(-13) & 1.96(-10) & 0.98 & 1 \\[10pt]

\multicolumn{13}{c}{{\footnotesize Reflection Nebulae}}\\[10pt]

NGC~7023 I\tablenotemark{f}        
                           & 21 01 31.90 & $+$68 10 22.12 & ISO-SWS~01(4) & 20700801 &  9  & 1.58(-13) & 9.61(-14) & 8.51(-14) & 3.72(-14) & 3.46(-11) & 0.38 & 22 \\

NGC~2023\tablenotemark{f}          
                           & 05 41 38.29 & $-$02 16 32.59 & ISO-SWS~01(2) & 65602309 &  10  & 8.28(-14) & 5.64(-14) & 4.67(-14) & 2.43(-14) & 3.04(-12) & -    &23 \\[10pt]

\multicolumn{13}{c}{{\footnotesize Embedded protostars}}\\[10pt]

NGC7538 IRS1               & 23 13 45.27 & $+$61 28 09.98 & ISO-SWS~01(3) & 38501842 & 11  & 2.17(-13) & 4.01(-12) & 4.06(-12) & 5.98(-12) & 1.80(-10) & -    &24 \\

MONR2 IRS2                 & 06 07 45.79 & $-$06 22 50.02 & ISO-SWS~01(1) & 71102004 &  -  & 4.82(-13) & 2.77(-12) & 2.60(-12) & 1.97(-12) & - & -    &\\[10pt]

\multicolumn{13}{c}{{\footnotesize Interstellar Medium line-of-sights}}\\[10pt]

Galatic Center SgrA*       & 17 45 39.97 & $-$29 00 28.70 & ISO-SWS~01(4) & 09401801 & 12   & 9.69(-14) & 4.80(-12) &           & 2.34(-11) & - &  - &  \\

Galatic Center Ring NE     & 17 45 41.80 & $-$28 59 50.50 & ISO-SWS~01(4) & 09500203 & 13   & 8.28(-14) & 1.08(-13) &           & 4.91(-13) & - &  - &   \\[20pt]

  &   &  & & &   & W/m$^2$/sr& W/m$^2$/sr & W/m$^2$/sr & W/m$^2$/sr & W/m$^2$/sr & & \\[10pt]

Rho-Oph                    & 16 25 41.09 & $-$24 06 46.90 & ISO-CAM~04    & 09202119 &  14 & 3.74(-07) & 1.15(-07) & 3.98(-07) & 1.66(-07) & 2.61(-05) & -    &14\\

SMC B1~\#~1 cloud          & 00 45 32.50 & $-$73 18 16.30 & ISO-CAM~04    & 23200127 &  15 & 2.29(-08) & 7.54(-09) & 6.50(-09) & 1.18(-09) & $<$0.46(-06) & -    &15\\*

                           &             &                &               & 68602088 & & & &  \\

\multicolumn{3}{l}{Milky Way - ISM (average)}      & ISO-PHOT-S & - &  16  & 3.72(-07) & 2.17(-12) & 2.61(-07) & - & 32.30(-06) & -    &15\\

\multicolumn{3}{l}{NGC891 (average within 144\arcsec from center)}   & ISO-PHOT-S & - &  16  & 3.72(-07) & 2.17(-12) & 2.61(-07) & - & 13.00(-06) & -    &16\\

\multicolumn{3}{l}{DISM 1 \hspace{1cm} 44\degr $\leq$ l $\leq$ 44\degr 40\arcmin, -0\degr 40\arcmin $\leq$ b $\leq$ 0\degr}
                                                             & IRTS           & -        & 17  & 1.31(-07) & 1.51(-07) & 2.44(-07) & - & 23.00(-06) & -    &17 \\
	                     
\multicolumn{3}{l}{DISM 2 \hspace{1cm} 50\degr $\leq$ l $\leq$ 53\degr, 1\degr $\leq$ b $<$ 2\degr} 
                                                             & IRTS           & -        & 17  & 5.14(-08) & 8.23(-08) & 1.35(-07) & - & 10.00(-06) & -    &17 \\
	                     
\multicolumn{3}{l}{DISM 3 \hspace{1cm} 51\degr $\leq$ l $\leq$ 54\degr, 2\degr $\leq$ b $\leq$ 3\degr} 
                                                             & IRTS           & -        & 17  & 2.00(-08) & 1.90(-08) & 4.81(-08) & - & 6.50(-06) & -    &17 \\

\enddata

\tablenotetext{a}{Units of $\alpha$ are hours,
    minutes, and seconds, and units of $\delta$ are degrees, arc
    minutes, and arc seconds.}

\tablenotetext{b}{SWS observing mode used
    \citep[see][]{deGraauw96}. Numbers in brackets correspond to
    the scanning speed.}

\tablenotetext{c}{Each ISO observation is given
    a unique TDT (Target Dedicated Time) number.}

\tablenotetext{d}{See text for details. Uncertainties on the MIR fluxes less than 20\%}

\tablenotetext{e}{FIR Scaling factor, see text for details.}

\tablenotetext{f}{MIR/FIR ratio influenced by beam effects, confusion with other sources or mispointings 
    \citep{Peeters02b, Martin:radio:02}.}

\tablenotetext{g}{In units of W/m$^2$/sr.}

\tablenotetext{h}{The contribution of IRC2 and BN is estimated on
    4(4)L$_\odot$ \citep{Thronson:86} and subtracted from the observed
    FIR luminosity.}

\tablenotetext{i}{Water ice absorption (6 \mum) present \citep{Peeters02b}.}

\tablerefs{
  1~:~\citet{Peeters:cataloog:01}; 
  2~:~\citet{Sturm00}
  3~:~\citet{Cesarsky:sileminorion:00};
  4~:~\citet{Verstraete01};
  5~:~\citet{Simpson:orionmsx:98};
  6~:~\citet{Roelfsema:pahs:96}; 
  7~:~\citet{Verstraete:m17:96};   
  8~:~\citet{Henning:98};
  9~:~\citet{Moutou:leshouches:99};
  10~:~\citet{Moutou:parijs:99};
  11~:~\citet{Gerakines99};
  12~:~\citet{Lutz:96}
  13~:~\citet{Lutz:99}
  14~:~\citet{Boulanger:oph:96} 
  15~:~\citet{Kahanpaa:ism:03};
  16~:~\citet{Reach:pahsinsmc:00};
  17~:~\citet{Onaka:dism:96};    
  18~:~\citet{Vermeij:pahs:01};
  19~:~\citet{Werner76};
  20~:~\citet{Thronson:84};
  21~:~\citet{Meixner:m17:92};
  22~:~\citet{Casey:91};
  23~:~\citet{Steiman-Cameron:97};
  24~:~\citet{Thronson:79} }

\end{deluxetable}


\clearpage

\subsection{ISO galaxy sample}
\label{sample_extra_galactic}

The MIR galaxy spectra presented here have been drawn from our
database of some 250 ISO galaxy spectra, described in
\citet{Spoon:silenpahs:01}. The sample comprises normal galaxies,
starburst galaxies, Seyfert galaxies, QSOs, LIRGs, ULIRGs and Hyper
Luminuous IR Galaxies (HyLIRGs).  Depending on the size of the
aperture used and the distance to the source, the spectra probe
physical sizes ranging from 73\,pc (4.5$\arcsec$) for the nearest
source (Cen\,A; D=3.5\,Mpc) and the entire disk (assuming R$_{\rm
disk}$=10\,kpc) for galaxies beyond 170\,Mpc. The MIR spectra are
supplemented with IR photometry from the IRAS Faint Source Catalog
(FSC). Given a beam size of $\sim$60$\arcsec$, the physical sizes
probed by IRAS range from 1\,kpc for the nearest source and the entire
disk for galaxies beyond 73\,Mpc. For details on the data reduction,
see \citet{Spoon:silenpahs:01}.

\section{The spectral characteristics}
\label{spchar}

In this section, we focuss on the spectral characteristics of our
sample sources in a qualititative way. A quantitative description is
given in Sect.~\ref{diagnostictools} and~\ref{discussion}.

\subsection{\HII regions and ISM}
\label{spchar_mw}

\begin{figure}[t] 
\epsscale{0.9}
\plotone{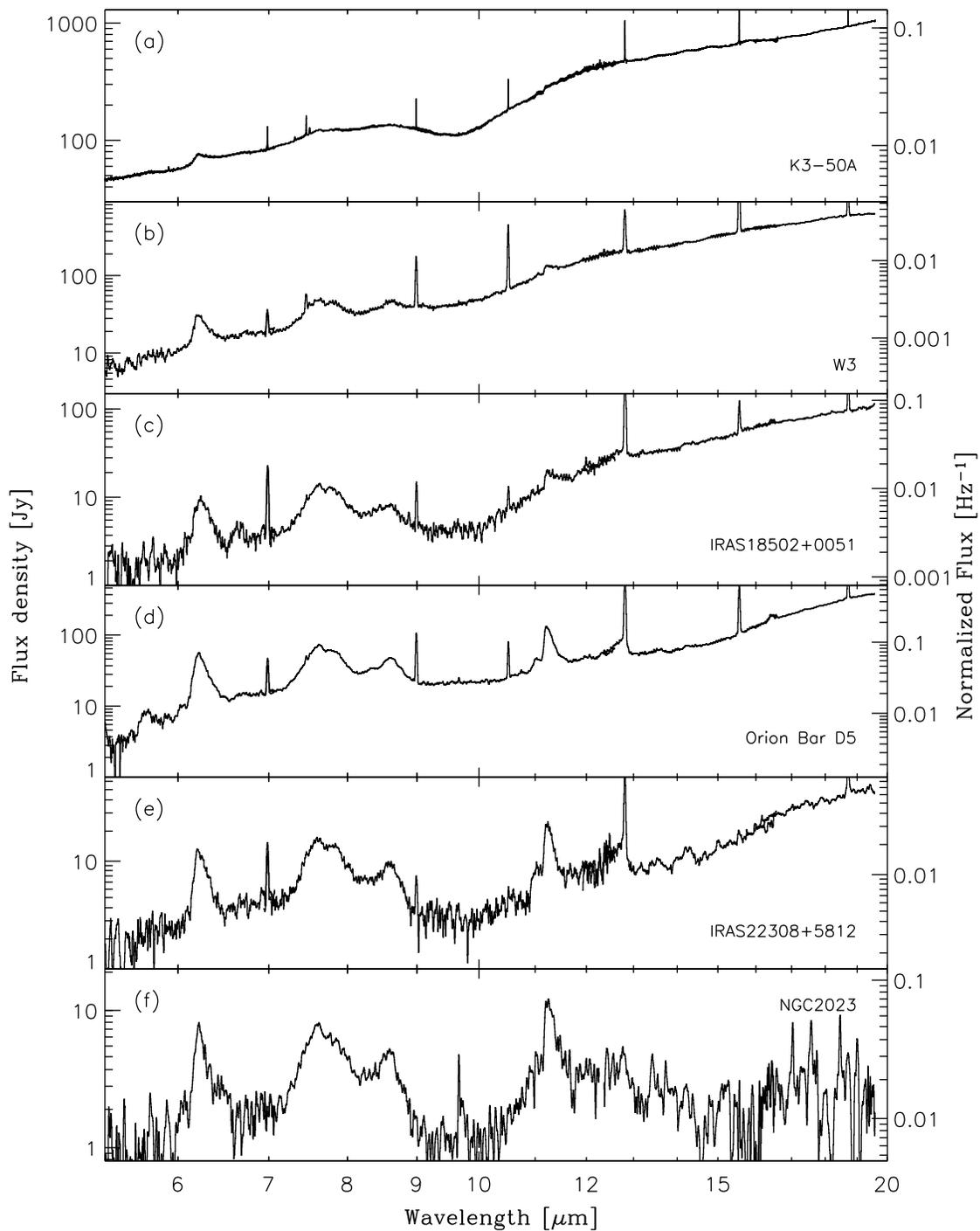}
\caption{Spectral variation in the MIR characteristics of \HII regions
\citep[for 2--200\,$\mu$m spectra, see][]{Peeters:cataloog:01}. As a reference,
the RN NGC\,2023 is shown in the bottom panel. The units on the right
vertical axis represent the FIR normalized flux. }
\label{zoo}
\end{figure}

Fig.~\ref{zoo} (b -- e) shows the MIR spectra of a few \HII
regions, carefully selected to span the range from highly embedded
\HII regions (e.g. W3) to optically visible \HII regions
(e.g. Orion). These spectra are characterized by a
strong rising dust continuum due to thermal dust emission,
corresponding to dust temperatures of $\sim$ 60--70 K. Many \HII
regions also show strong continuum emission at $\lambda < 12$\mum,
whose origin is unclear. Possibly, this continuum is due to a small
fraction of dust inside the \HII region, heated to high temperatures
by resonantly scattered Ly$\alpha$ radiation. Alternatively, these are
larger ($\sim$ 500 C-atoms) PAH-like structures stochastically heated
by a single or multi-photon event.  On top of the dust continuum,
there is a series of fine-structure lines and hydrogen recombination
lines.  In addition, these spectra exhibit prominent PAH emission
features, often silicate absorption and in some cases absorption bands
due to molecular ice species (CO$_2$, H$_2$O).

In contrast, RNe exhibit a much weaker dust continuum indicating lower
dust temperatures, strong PAH emission and no fine-structure lines or
recombination lines (Fig.~\ref{zoo}~f).  A peculiar spectrum is that
of the heavily obscured \HII region K3--50A, showing all
characteristics of \HII regions except for the dust continuum which is
more similar to that of massive protostars (Fig.~\ref{zoo}~a). This
source also exhibits a multitude of ice absorption features
\citep{Peeters:cataloog:01}.

\clearpage

\clearpage

Each of these emission/absorption components characteristic of \HII
regions is influenced by the local physical conditions and hence,
although their global characteristics are very similar, the individual
sources display great diversity in their spectral properties
\citep[for this sample see][]{Roelfsema:pahs:96, Verstraete:m17:96,
Verstraete01, Hony:oops:01, Peeters02b, Peeters:cataloog:01, paperii,
Vermeij:pahs:01, vanDiedenhoven:chvscc:03}.  Indeed, large variations
are present in the relative strength of the PAH emission bands and the
dust continuum at various wavelengths (Fig.~\ref{zoo}). While the PAH
features are relatively weak compared to the continua in the spectra
presented in the top panels, the opposite is true for the spectra in
the lower panels, indicating a smooth trend among the \HII regions,
ultimately going to that of RNe.  However, the strength of the thermal
emission of dust below 20 \mum\, is highly influenced by the dust
temperature and the amount of dust. Indeed, when normalized on the
far-infrared (FIR) flux (Fig.~\ref{zoo} : right y-axis), the sequence
of sources going from low to high relative strength of PAH over FIR
dust continuum ratio is different, but, here as well, going toward the
high ratios of RNe. Both PAH to continuum ratios (i.e. PAH emission
with respect to both MIR and FIR continua) vary from highly embedded
ultra-compact \HII regions (e.g. W3) toward extended optically visible
\HII regions (e.g. Orion) and ultimately toward RNe (e.g. NGC\,2023).

Similarly, the ratio of PAH over hot continuum (e.g. around $\sim$ 6
\mum) changes, albeit over a small range. The latter ratio can be
influenced by the possible presence of a broad emission plateau
underneath the 6.2, 7.7 and 8.6 \mum\, PAH bands of variable strength,
sometimes rivaling the strength of the dust continuum at this
wavelength. This emission plateau starts longwards of $\sim$ 6 \mum\,
and extends up to $\sim$ 9 \mum.

The profile and position of the PAH bands in \HII regions and RNe show
very little variation from source to source \citep{Peeters02b,
vanDiedenhoven:chvscc:03}, although spatially within a source
differences have been observed \citep[][Joblin~et~al., in
prep.]{Bregman:04}. The relative strength of the PAH bands and 
the fraction of the total PAH flux emitted in each individual band,
varies significantly from source to source and spatially within a
source.  This is very clear when e.g. comparing the 6.2/11.2 PAH ratio
for IRAS\,18502 and the Orion Bar \citep[for a detailed discussion, see
e.g.][]{Verstraete:m17:96,Joblin:isobeyondthepeaks:00, Hony:oops:01,
Vermeij:pahs:01, Peeters02b, Peeters:review:04}. \\

Clearly, the detailed spectral characteristics of \HII regions varies
and hence no ``typical'' spectrum exists when considering relative
strengths of different emission/absorption components. Since this is
partly due to the inclusion of emission of the surrounding Photo
Dissociation Region (PDR) and molecular cloud in the aperture, many
authors \citep[e.g.][]{Laurent00, Sturm00} have considered the
spectrum of starforming regions where continuum emission at 15 \mum\,
is dominant (e.g. M\,17) as the typical MIR spectrum for the \HII
region itself, and a reflection nebula (e.g. NGC\,7023) as the typical
MIR spectrum of a PDR. Indeed, the PDRs associated with compact \HII
regions, such as Orion, dominate the PAH emission from the region as a
whole \citep{Chrysostomou92,Giard92,Giard94,Graham93,Sellgren90,
Tielens:anatomyorionbar:93, Verstraete:m17:96, Contursi:N4:98}. The
spectra of \HII regions measured within a large beam are then
considered to be a combination of these two typical spectra. However,
some highly embedded \HII regions have a distinct dust continuum
(Fig.~\ref{zoo}) and therefore are not taken into account in this
decomposition. Sect.~\ref{diagnostictools} discusses to which extent
this decomposition is valid.

\subsection{Galaxy sample}
\label{spchar_galaxies}

\clearpage

\begin{figure}[!t]
\begin{center}
\resizebox{\hsize}{!}{{\includegraphics{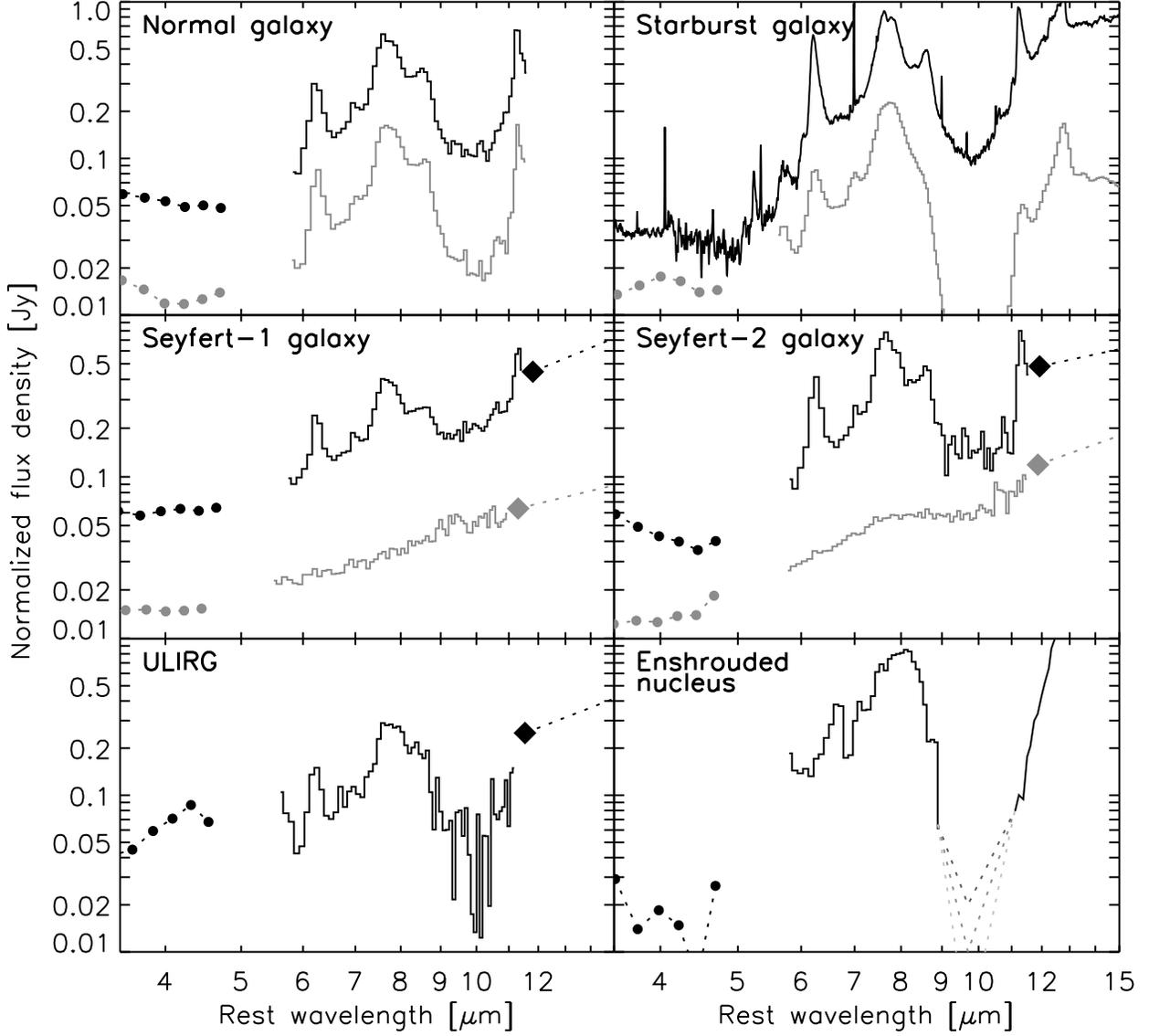}}}
\end{center}
\caption{MIR galaxy spectra. Dots represent ISO--PHT--SS spectra
rebinned to a lower resolution, diamonds the IRAS 12\,$\mu$m flux
shown only when no aperture flux mismatches are seen between ISO and
IRAS (see Sect.~\ref{plotje_zonder_naam}). For ease of presentation,
the spectra are divided by an arbitrary factor given after each
source. {\bf Top-left panel:} The 'normal' galaxies NGC\,4102 (3, {\it
black}) and NGC\,3620 (10, {\it grey}).  {\bf Top-right panel:} The
nuclei of the starburst galaxies M\,82 (45, {\it black}) and NGC\,4945
(25, {\it grey}).  {\bf Mid-left panel:} The Sf1 galaxies NGC\,7469
(3, {\it black}) and I\,Zw\,1 (8, {\it grey}).  {\bf Mid-right panel:}
The Sf2 galaxies NGC\,5953 (1.1, {\it black}) and PKS\,2048--57 (9,
{\it grey}).  {\bf Bottom-left panel:} The ultra-luminous IR galaxy
UGC\,5101 (1). {\bf Bottom-right panel:} The heavily obscured nucleus
of NGC\,4418 (1.2). }
\label{galaxy_zoo}
\end{figure}

\clearpage

The MIR spectra of normal and starburst galaxies reveal the same
spectral features that constitute the spectra of Galactic star forming
regions. Invariably, the 6--12\,$\mu$m range is dominated by the PAH
bands at 6.2, 7.7, 8.6 and 11.2\,$\mu$m
(Fig.~\ref{galaxy_zoo}). Beyond 9\,$\mu$m, starburst galaxies show the
additional presence of a warm dust continuum, which is weak or absent
in normal galaxies.  High resolution spectra also reveal many strong
emission lines.

The detailed characteristics of the IR spectra depend on the
properties of the galaxy in a way which is not completely
understood. In particular, some starburst galaxies also show evidence
for the presence of cold, absorbing dust in the nuclear region. A
clear example of this is NGC\,4945 (top-right, Fig~\ref{galaxy_zoo}),
as can be deduced from the weakness of the continuum and PAH features
closest to the center of the 9.7\,$\mu$m silicate absorption feature.

Galaxies optically classified as Seyferts contain an AGN -- a massive
central black hole, surrounded by an X-ray emitting accretion
disk. The disk is believed to be surrounded by a thick molecular
torus, the orientation of which determines the Seyfert
subtype. Sf1s offer a direct line of sight to the
accretion disk and the hot dust at the inner face of the molecular
torus. For Sf2s, this direct view is blocked by the torus. Gas
gas above the plane of the torus, highly ionized by the X-rays from
the accretion disk, serves to betray the presence of the AGN.  The MIR
spectra of Seyferts are quite diverse, with some spectra
bearing close resemblance to those of starburst galaxies, while other
spectra are dominated by a hot dust power law spectrum (middle panels,
Fig.~\ref{galaxy_zoo}).  Although the average Sf1
spectrum is continuum-dominated and the average Sf2 galaxy
spectrum is not \citep{Clavel00}, our ISO spectral library contains
many examples of Seyfert spectra showing the properties of the
converse subtype (middle panels, Fig.~\ref{galaxy_zoo}). Evidently,
the overall MIR spectral appearance cannot be just the result of the
orientation of the AGN torus, but likely also depends on the
properties of the host galaxy (i.e. inclination, degree of starburst
activity) as well as the power of the central engine.  The spatial
scale of the observations is another factor determining the MIR
spectral appearance of AGNs. Observations of nearby PAH dominated
Sf2s, like Circinus and Cen\,A, have shown the spectrum
of the inner 70--150\,pc to be dramatically different from the
spectrum of the entire galaxy, revealing an underlying,
silicate-absorbed, hot dust continuum, likely associated with the
AGN \citep{Laurent00}.

A very different MIR spectrum is detected towards the nucleus of the
luminous IR galaxy NGC\,4418 (bottom-right,
Fig.~\ref{galaxy_zoo}). The spectrum is characterized by a prominent
broad peak, centered at 8\,$\mu$m, too wide and too far displaced to
be attributed to 7.7\,$\mu$m PAH band. Instead, the spectrum appears
to be the result of the absorption of a featureless continuum by ices
and silicates, both shortward and longward of 8\,$\mu$m, reminiscent
of the spectrum of a deeply embedded protostar
\citep{Spoon:ngc4418:01}. MIR imaging has further revealed the source
of the absorbed MIR continuum to be extremely compact
\citep[$<$80\,pc;][]{Evans03} and to be responsible for most of the IR
luminosity. In the absence of any spectral line information, the
nature of the nuclear energy source is unclear and may either be
extremely dense star formation or AGN activity or a combination of
both \citep[e.g.][]{Spoon:ngc4418:01, Evans03}. Several other galaxies
have been found with spectra similar to NGC\,4418, most of them are
ULIRGs \citep{Spoon:silenpahs:01}.

\begin{figure}[!t]
\epsscale{0.8}
\plotone{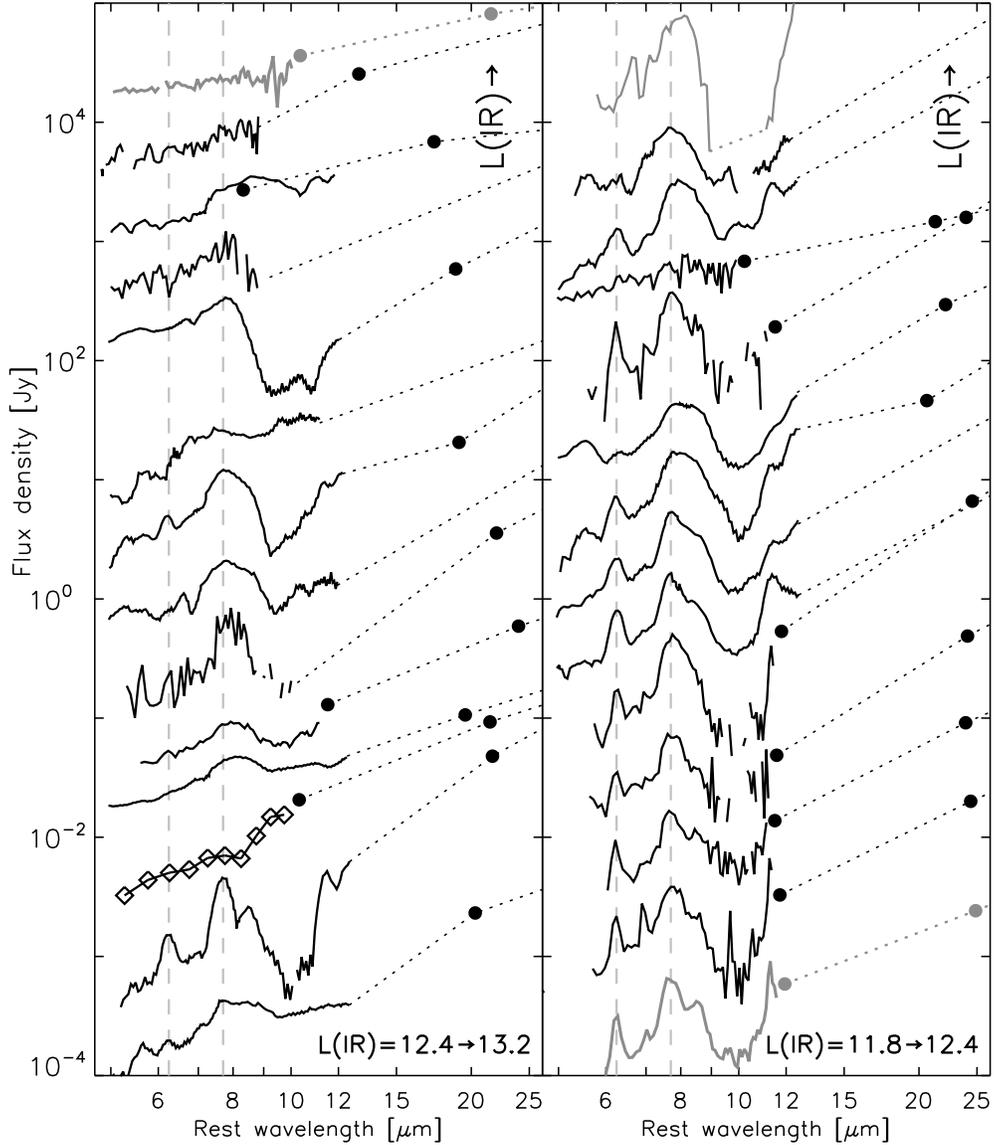}
\caption{HyLIRG and ULIRG spectra sorted by increasing IR luminosity.
{\it Vertical lines} indicate the central wavelengths of the
6.2\,$\mu$m and 7.7\,$\mu$m PAH features and {\it solid circles} the
IRAS 12 and 25 \mum\, fluxes (when not an upperlimit).  The spectra
shown in {\it grey} serve as template spectra for AGNs ({\bf
top-left}), starbursts ({\bf bottom-right}) and deeply enshrouded
nuclei ({\bf top-right}). The spectra in the {\bf left panel} are,
from top to bottom: 3C\,273, 15307+3252, 09104+4109, 00397--1312,
00183--7111, 23529--2119, 03538--6432, 23515-2917, 03158+4227,
Mrk\,231, 00275--2859, Mrk\,1014, 03521+0028 and 22192--3211. In the
{\bf right panel}: NGC\,4418, 17463+5806, 03000--2719, 23060+0505,
17208--0014, 00188--0856, 04384--4848, 02113--2937, 18030+0705,
Arp\,220, Mrk\,273, 23128--5919, NGC\,6240 and NGC\,7552.}
\label{ulirg_sequence}
\end{figure}

The MIR spectra of ULIRGs are more complicated than the spectra of
other galaxies due to the presence of copious amounts of dust in their
nuclear regions, leading to strong absorption features which may
spectrally distort any emission component. Fig.~\ref{ulirg_sequence}
shows a compilation of available HyLIRG and ULIRG spectra, sorted by
IR luminosity.  While most low-luminosity ULIRGs are dominated by PAH
features, the spectra at the high-luminosity end bear close
resemblance to AGN hot dust continua, showing little or no sign of PAH
features. At intermediate luminosities some ULIRG spectra are PAH
dominated, while others show signs of the presence of an
NGC\,4418-like broad 8\,$\mu$m peak, indicating the presence of deeply
enshrouded power sources. In addition, some of these ULIRGs show
absorptions of water ice \citep{Spoon:silenpahs:01}. A typical, PAH
dominated, ULIRG spectrum is further displayed in the lower-left panel
of Fig.~\ref{galaxy_zoo}.

\clearpage

\clearpage

\section{Diagnostic tools to distinguish AGNs and starburts}
\label{diagnostictools}

As AGNs and star forming regions have distinct spectral
characteristics, their relative contribution to observed MIR spectra
can be determined in various ways. In this section, we focus on three
different diagnostic tools, all based on the strength of PAH bands.

\subsection{A MIR/FIR diagnostic}
\label{plotje_zonder_naam}

Indicative of regions predominantly powered by star formation are the
strong PAH bands. As a measure of the strength of these PAH bands, we
prefer the 6.2\,$\mu$m band since: 1) unlike the 3.3 and 11.2 $\mu$m
bands, it could be observed by all three spectrometers on board ISO;
2) it is situated well outside the silicate absorption band and is the
least influenced by extinction and 3) unlike the 7.7\,$\mu$m feature,
it cannot be confused with a NGC\,4418-like absorbed-continuum peak
\citep{Spoon:ngc4418:01, Spoon:silenpahs:01}. A clear line of sight
towards AGN-heated hot dust may be recognized by the strong
6.2\,$\mu$m continuum associated with a hot dust power law
spectrum. Since hot dust predominantly radiates at MIR wavelengths,
the ratio of 6.2\,$\mu$m continuum over FIR flux will be high, far
higher than for both star forming regions and AGNs without a clear
line of sight to the AGN-heated hot dust.  We therefore normalize our
star formation indicator and our AGN hot dust indicator on the FIR
flux (referred to as 6.2PAH/FIR and 6.2cont/FIR). A
clear disadvantage of this diagnostic is the often different apertures
of the MIR and FIR instruments.

The continuum is estimated by drawing a polynomial of order 2
(i.e. straight line) through points at $\sim$5.9 and $\sim$6.6
\mum. Integrating the flux underneath it and above it in the range
from 6.0 to 6.5 \mum\, gives respectively the strength of the hot
continuum and the strength of the 6.2\,$\mu$m PAH feature. If a source
exhibit water ice absorption (see Table~\ref{log} for Galactic sources
and \citet{Spoon:silenpahs:01} for the galaxies), a local spline
continuum is determined along the wing of the water ice band to
estimate the 6.2\,$\mu$m PAH flux for Galactic sources while for
extragalactic sources the 6.2\,$\mu$m PAH flux is recovered by
defining a continuum at the base of the 6.2\,$\mu$m PAH feature. In
order to obtain continuum strength corrected for ice extinction, the
6\,$\mu$m continuum strength is estimated by integrating a spline
fixed just outside the ice absorption feature for these Galactic
sources and by integrating the model continua derived in
\citet{Spoon:silenpahs:01} for galaxies. The spectra of some galaxies
show no evidence for 6.2\,$\mu$m PAH emission. For these galaxies,
upper limits for the 6.2\,$\mu$m PAH flux are obtained by integrating
a Gaussian with a peak flux of three times the rms noise and a FWHM of
0.185\,$\mu$m.

For most galaxies, the FIR flux is computed from the 60 and
100\,$\mu$m IRAS Faint Source Catalog (FSC) fluxes, applying the
formula for F(40--500\,$\mu$m) as given in \citet{Sanders96}. We
assumed the correction factor C equaling 1.0 for IRAS60/100$>$1.0 and
linearly increasing to 1.6 for IRAS60/100=0.3. The linear part of the
correction factor is based on the galaxies Mrk1116, NGC3583, NGC4194
and NGC6090, for which we were able to perform a direct integration of
the FIR SED.  For galaxies with redshifts in excess of z=0.1,
the FSC fluxes are a progressively worse probe of the rest frame 60
and 100\,$\mu$m fluxes. For these galaxies and galaxies lacking a
100\,$\mu$m FSC detection, we have supplemented the FIR SED with
literature data and determined F(40--500\,$\mu$m) by direct
integration of the obtained 40--500\,$\mu$m rest frame SED.  For those
galaxies for which additional FIR and submm literature data are not
available, we looked for an observed SED that best matches the 5-60
\mum\, spectral properties and used its FIR and submm data beyond 60
\mum\, to compute the F(40--500\,$\mu$m) by direct integration of the
40--500\,$\mu$m rest frame SED.  If no reliable match was found, we
completed the SED with both a warm observed SED (e.g. 05189-2424) and
a cold observed SED (e.g. IRAS19458+0944), resulting in two bracketing
FIR fluxes.  For QSOs we also considered power laws to provide the
upper limits to the FIR flux.  As pointed out in
Sect.\,\ref{spchar_galaxies}, ISO beams probe smaller physical scales
than the IRAS beam. This mismatch becomes an issue when relating MIR
to FIR fluxes for nearby galaxies for which the IR luminous regions do
not completely fit within the smaller of the two apertures.  We
therefore screened our sample against spectra showing clear mismatches
between the IRAS 12\,$\mu$m flux and the ISO 11--12\,$\mu$m continuum
flux. Our sample should thus only consist of galaxies whose MIR
emitting region fits entirely within the ISO aperture. For the FIR
flux we will assume that it is dominated by the same (circum)nuclear
star forming regions which give rise to the MIR continuum.
Subsequently, the sample is screened against any spectrum for which
the 6.0--6.5\,$\mu$m continuum is so noisy that the integrated
6.2\,$\mu$m flux would have to be replaced by an upper limit.  Our
final galaxy sample consists of 69 AGNs (7 QSO, 31 Sf1s and 31 Sf2s),
22 starburst and \HII-type galaxies, 8 normal galaxies, 49 ULIRGs, 2
HyLIRGs and 3 IR galaxies with deeply obscured nuclei.

For the compact \HII regions, the FIR flux is derived by fitting a
modified blackbody to the IRAS PSC fluxes
\citep{Peeters:cataloog:01}. However, ISO-SWS and IRAS have different
beams (20$\arcsec$ versus 90$\arcsec$).  This may seriously affect the
derived ratio of the MIR/FIR ratios if the source shows structure on a
scale larger than the SWS beam. Therefore, we derive a scaling factor
by matching the observed LWS flux to the SWS flux at 45 \mum.  Since
the IRAS PSC fluxes -- used to derive FIR fluxes -- nicely correspond
to the flux seen by ISO-LWS in the IRAS bands at 60 and 100 \mum\, for
most sources \citep{Peeters:cataloog:01}, this scaling factor is a
good correction. The derived FIR flux may still be incorrect for
extended and complex sources. Since they nevertheless fit in with the
overall envelope, we do show them in Fig.~\ref{contpahfir_galactic}
in grey and indicate them in Table~\ref{log}.  For the extended
molecular cloud M17--North, 30Dor and the RN NGC\,7023, we integrated
the LWS flux (45-200 \mum) and scaled it with the above defined
scaling factor.  For a few sources, higher spatial resolution data
(e.g. KAO with a beam of 35--60$\arcsec$) exist and we have used those
to derive a FIR flux.  M17 IRAM pos. 1 and 8 correspond to positions I
and II in \citet{Meixner:m17:92} and hence their FIR flux is derived
from the incident UV flux, G$_0$ given by these authors and corrected
for the beam difference. The FIR flux for Orion (MSX) and NGC7538 IRS1
is the FIR flux obtained with KAO observations, corrected for the
difference in the beams. The ISO-SWS position in NGC2023 is closest to
position H4 of \citet{Steiman-Cameron:97}. We obtained its FIR from
the FIR given for position P4 scaled by the ratio of the 64 \mum\,
continuum flux of position H4 to P4 \citep{Steiman-Cameron:97}. The
FIR of $\rho$~Oph, SMC B\#1, DISM 1,2,3 and Milky Way (average) is
estimated from their 100 \mum\, continuum fluxes (135 \mum\, continuum
flux for SMC B\#1) assuming a similar SED as that of cirrus
\citep{Boulanger:98}. We estimated the FIR for the central region of
NGC891 from the observed UIR flux (in this central region) and the
UIR/FIR ratio for the whole galaxy \citep{Mattila:99}. For the Orion
Bar the FIR is taken from \citet{Werner76}.
 
The derived fluxes and the FIR references for the Galactic sample and
the extragalactic sample are given in, respectively, Table~\ref{log}
and Spoon et al. (in prep.).

\subsubsection{\HII regions and ISM}
\label{mw_plotje_zonder_naam}

\begin{figure}[!t]
\begin{center}
\resizebox{\hsize}{!}{{\includegraphics{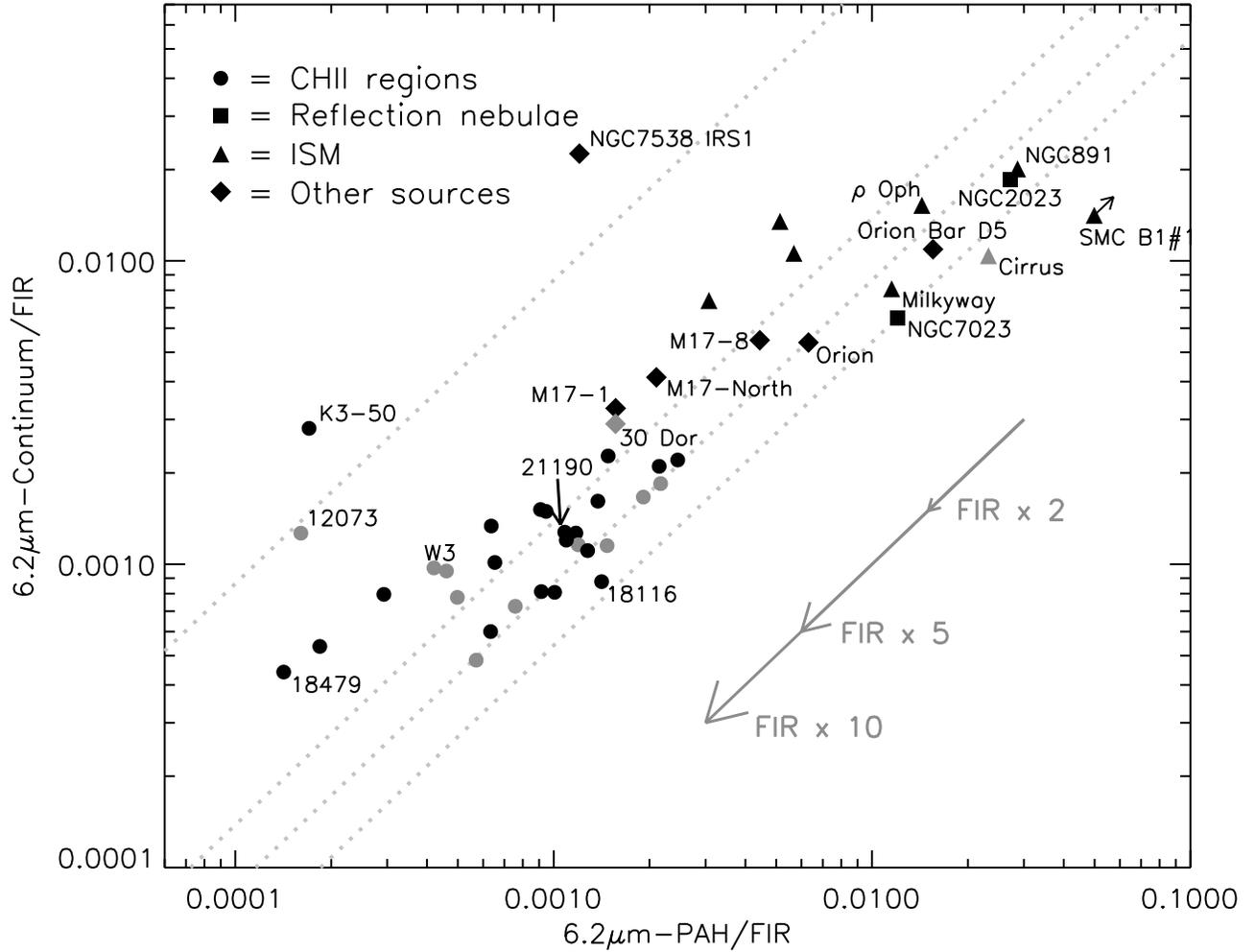}}}
\end{center}
\caption{A MIR/FIR diagnostic diagram for the Galactic sample. The
influence of an uncertain FIR flux is indicated by the arrows for
respectively a factor of 2, 5 and 10. The {\it dotted lines}
correspond to a 6.2PAH/6.2cont ratio of 0.116, 0.72, 1.16 and 1.85 from
top to bottom.  The sources possibly suffering from aperture effects
are plotted in {\it grey}.}
\label{contpahfir_galactic}
\end{figure}

As discussed in Sect.~\ref{spchar_mw}, the 6.2PAH/FIR and the
6.2cont/FIR vary clearly within our sample of \HII regions and this
sequence extends up to the RNe and the (diffuse) ISM lines of sight
(Fig.~\ref{contpahfir_galactic}). Most sources are well localized in
this diagram and are positioned along a strip going from the lower
left towards the upper right, further referred to as the `fundamental
line'. A clear segregation is present with embedded compact \HII
regions situated on the lower left side, while the RNe and the
(diffuse) ISM are on the top right side. In addition, situated in the
middle-right are the exposed PDRs, such as M\,17 (with pointings to
the \HII region, M\,17--1, and the molecular clouds, M17--8 and
M17--North) and Orion. This segregation with object type suggests that
the underlying cause may be a variation in the physical/chemical
properties of the PAHs going from compact \HII regions to exposed PDRs
and the diffuse ISM, e.g. with G$_0$/n$_e$
\citep{Bakes01a,Bakes01b}. Whether the segregation within the sample
of \HII regions is dominated by an evolutionary sequence --- from
ultra-compact to compact and then classical \HII regions --- is less
clear. The 6.2PAH/FIR ratio correlates weakly with both the electron
density and the size of the \HII region \citep[taken from][]{paperii,
  Martin:radio:02}. In contrast, the 6.2cont/FIR does not show a clear
dependence on either the electron density or the size of the \HII
region.  The latter also show large scatter in
Fig.~\ref{contpahfir_galactic}, where many \HII regions lie above the
fundamental line. Likely, this reflects the contribution of hot dust
inside the \HII region.  We note that metallicity does not seem to
influence the location in this diagram. In
particular, the SMC B1\#1 and 30\,Dor --- with metallicities 0.3 and
0.1 of solar --- lie close to their Galactic counterparts.

\clearpage

\clearpage

To guide the eye, we have drawn lines of constant 6.2PAH/6.2cont in
Fig.~\ref{contpahfir_galactic}. This ratio varies within the sample,
ranging from 0.06 to 3.5 with an average of 1.0$\pm$0.6 (0.93$\pm$0.47
for only the \HII regions). There seems to be a systematic trend: \HII
regions tend to lie above the bottom two lines while RNe and the diffuse
ISM tend to lie between the bottom two lines.  Three notable
outliers are the Orion Bar, IRAS~12073 and K3--50A. The Orion Bar is
located with the RNe suggesting a PDR dominated spectrum
\citep[cf.][]{Tielens:anatomyorionbar:93}. IRAS~12073 has the lowest
6.2PAH/6.2cont ratio of the \HII regions due to its very strong hot
dust continuum. Indeed, IRAS~12073 is also the source with the largest
Lyman continuum luminosity amongst the \HII regions.  As already
pointed out, K3--50A has all the characteristics of \HII regions
except for its continuum which is more like that of massive protostars
(see Sect.~\ref{spchar_mw}) and therefore likely has a similar
6.2PAH/6.2cont ratio as the highly embedded, pre-main-sequence object
NGC\,7538\,IRS1 but shows the 6.2PAH/FIR characteristics of compact
\HII regions.

Based upon a comparison of \ion{H}{1} recombination line fluxes and
radio free-free fluxes, extinction in the MIR should be negligible for
our sample of \HII regions with $\tau_{\rm dust}(6.2) \leq 1$
\citep[][an A$_K$ of 2 would decrease the 6.2PAH/FIR and 6.2cont/FIR
with $\sim$20\%]{paperii,Martin:radio:02}.  Hence, we conclude that the
presence of a range in the 6.2PAH/FIR and 6.2cont/FIR ratios is real
and that the sequence spanned by our sample likely reflects an
``evolution'' with object type : from embedded compact \HII regions
towards exposed PDRs and (diffuse) ISM.

\subsubsection{The galaxies}
\label{gal_plotje_zonder_naam}

\clearpage

\begin{figure}[!t]
\epsscale{0.65}
\plotone{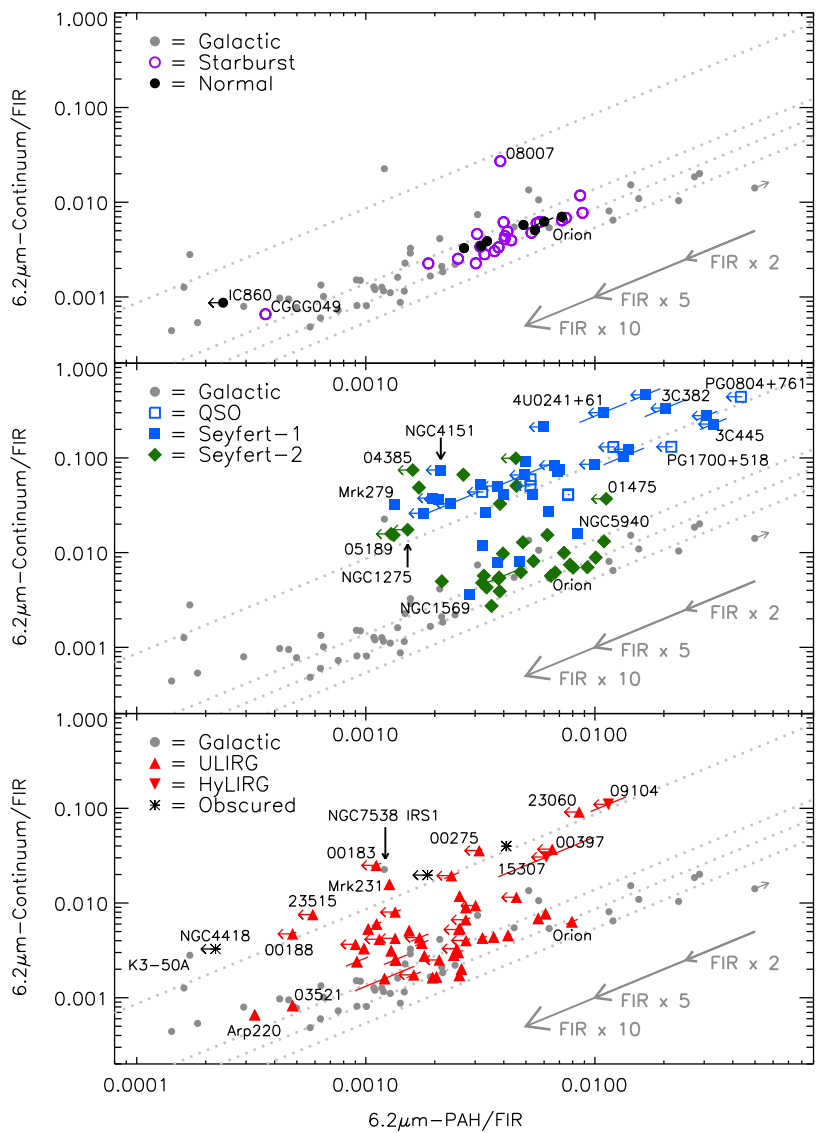}
\caption{MIR/FIR diagnostic diagram for different type of galaxies.
{\it Grey} circles correspond to Galactic sources. The
three parallel {\it dotted lines} and the arrows are as in
Fig.\,\ref{contpahfir_galactic}. }
\label{contpahfir_panels}
\end{figure}

\clearpage

The normal galaxies occupy a narrow strip close to the position of the
exposed PDR Orion (top panel, Fig.\,\ref{contpahfir_panels}). Note the
extreme position of IC\,860, also found by \citet{Lu:03}. ISO--LWS
observations have revealed this source to be very unusual in not
showing any of the typical FIR fine structure lines
\citep{Malhotra01}.  Also the ISO--PHT--S spectrum is unusual, as it
does not show 6.2\,$\mu$m PAH emission \citep[see also
e.g.][]{Lu:03}. The quality of the spectrum is, however, not good
enough to speculate on the nature of the MIR spectrum.

Most starburst galaxies are found close to the position of the normal
galaxies (top panel, Fig.\,\ref{contpahfir_panels}), with the
exception of IRAS\,08007--6600 and CGCG\,049--057. The ISO--PHT--S
spectrum of IRAS\,08007--6600 (classified by \cite{Veilleux87} as
\HII-type) looks very similar to AGN hot dust dominated spectra and
was therefore classified by \cite{Laureijs00} as AGN-dominated. The
spectrum of CGCG\,049--057 is PAH-dominated, but
has a very strong and cold FIR continuum.  Note that the starburst
galaxies (excluding the latter two) occupy a larger range along the
fundamental line than normal galaxies do, and look, on average, like
exposed PDRs with similar 6.2PAH/6.2cont ratios.

The distribution of AGNs seems to segregate into two groups (middle
panel, Fig.~\ref{contpahfir_panels}). One group (group A) is made up
mostly of Sf2s with IRAS S\,25/S\,60 ratios all below 0.25, while the
other group (group B) consists mostly of Sf1s and QSOs with IRAS
S\,25/S\,60 ratios all larger than 0.25. The galaxies in group A all
show 6.2PAH/FIR ratios in the same range as normal and starburst
galaxies. In contrast, most of the galaxies in group B only
have formal 6.2PAH/FIR upper limits, some of which are found at
6.2PAH/FIR ratios 2--3 times larger than the highest value found for
normal and starburst galaxies. These upper limits shift left and fit
in with the rest of their group (B) if the scaling by FIR flux is replaced
by a scaling by the total IR flux
\citep[8--1000\,$\mu$m;][]{Sanders96}. In line with the orientation
dependent unification scheme \citep{Antonucci93}, the segregation in
6.2cont/FIR between the two groups likely reflects the orientation of
the toroid structure around the nucleus which for the lower
group (A) may block our line of sight to the warm inner toroid,
responsible for the 6\,$\mu$m continuum.  Assuming this to be the
case, a separation of a factor 10 in 6.2cont/FIR may be interpreted as
an A(V)=50--150 (depending on the choice of extinction law), which
(assuming a normal gas to dust ratio N$_{\rm H}$=1.9$\times$10$^{21}$
A$_{\rm V}$) is equivalent to a column density of
10$^{23.0}$--10$^{23.5}$\,cm$^{-2}$. This range is in good agreement
with the mean Sf2 absorbing column, as measured directly from X-ray
data \citep{Risaliti02}, and with the results of \cite{Clavel00}.  The
presence of some Sf1s in groups A (NGC\,1569, Mrk\,789,
NGC\,5940 and NGC\,7469) would then imply that the intrinsic power of
the AGN in these galaxies must be small for the AGN hot continuum not
to dominate the MIR spectral appearance.  Alternatively, the AGN hot
continuum and the optical AGN diagnostic lines emanating from these
galaxies are obscured by patchy foreground dust clouds along the line
of sight in the host galaxy, but still allowing an optical
classification.  Conversely, the presence of some Sf2s in 
group B (NGC\,1068, NGC\,1275, Mrk\,463, IRAS\,04385--0828, NGC\,5506
and PKS\,2048--57) would then be explained by a direct line of sight,
at grazing angle, to the inner edge of the torus, or by the presence
of dust clouds above the plane of the torus, which are irradiated and
heated by the central source or a more powerful AGN.

\begin{figure}[!t]
\begin{center}
\resizebox{\hsize}{!}{{\includegraphics{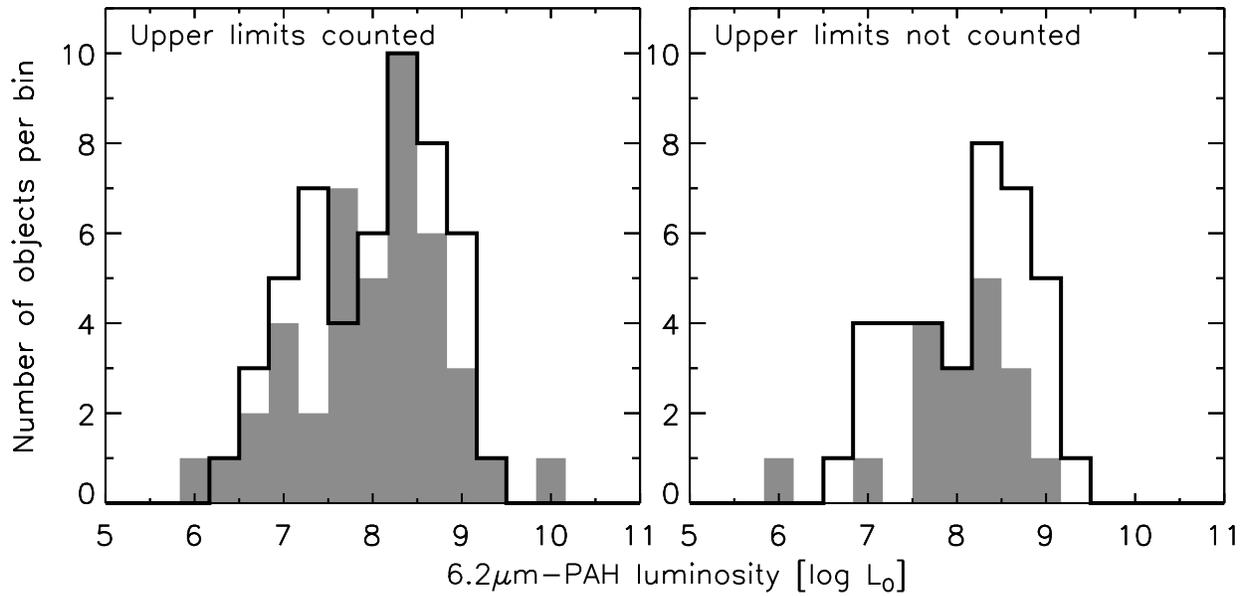}}}
\end{center}
\caption{Luminosity distribution of the 6.2\,$\mu$m PAH feature for
Sf1s ({\it filled grey}) and Sf2s ({\it black}).
The bin sizes are 0.33 dex. {\bf Left panel:} Upper limits on the 
PAH flux are counted as detections. {\bf Right panel:}
Upper limits on the PAH flux are excluded.}
\label{seyfert_pah_dist}
\end{figure}

One of the key predictions of the unification scheme is that the
properties of the Seyfert host galaxies are independent of the
orientation of the torus and, hence, independent of Seyfert type.
\cite{Clavel00} have tested this with the 7.7 \mum\, PAH luminosity
distributions for Sf1s and Sf2s and found that they were the same
for their sample. We have repeated this test for our sample of Seyfert
galaxies (43 Sf1s; 51 Sf2s). The resulting luminosity distributions
are shown in Fig.~\ref{seyfert_pah_dist}. If upper limits are counted
as detections, both Seyfert types have the same median 6.2 \mum\, PAH
luminosity of 10$^{8.1}$\,L$_{\odot}$. If upper limits are excluded,
the median values are still very similar, 10$^{8.2}$\,L$_{\odot}$ for
Sf1s and 10$^{8.3}$\,L$_{\odot}$ for Sf2s. Our results are therefore
consistent with those of \cite{Clavel00}.

\clearpage

\clearpage

Even more dusty are the nuclei of most ultra-luminous IR galaxies.
ULIRGs, HyLIRGs and galaxies with deeply obscured nuclei cover a large
portion of the diagram, reflecting their wide-ranging properties
(lower panel, Fig.\,\ref{contpahfir_panels}). Galaxies in the upper
right section (e.g. I\,00397--1312, I\,00275--2859, I\,09104+4109,
I\,15307+3252 and I\,23060+0505), all have IRAS S\,25/S\,60 ratios and
MIR spectra similar to AGNs in that same section and are likely
AGN-dominated. In contrast, some ULIRGs show spectra similar to Orion
and other bonafide star formation regions and are likely dominated by
star formation.  Many other ULIRGs occupy a space not covered by any
other (extra)galactic source and are situated to the left of both Seyfert
distributions.  The nuclei of these ULIRGs are therefore evolving from
completely embedded to less obscured nuclei.  Notable outliers are
Arp\,220 and I\,03521+0028, which are found among the C\HII regions,
close to the position of another outlier, IC\,860. Since the spectrum
of Arp\,220 is deficient in emission lines \citep{Spoon:arp220:04} --
characteristic of C\HII regions -- this source is likely shifted down
from the ULIRG domain, due to the presence of an additional strong FIR
component, which is not commonly present in other ULIRGs
\citep{Spoon:arp220:04}.  Given their close proximity to Arp\,220,
I\,03521+0028 and IC\,860 may also harbor similar FIR bright
components.

\clearpage

\begin{figure}[!t]
\begin{center}
\resizebox{\hsize}{!}{{\includegraphics{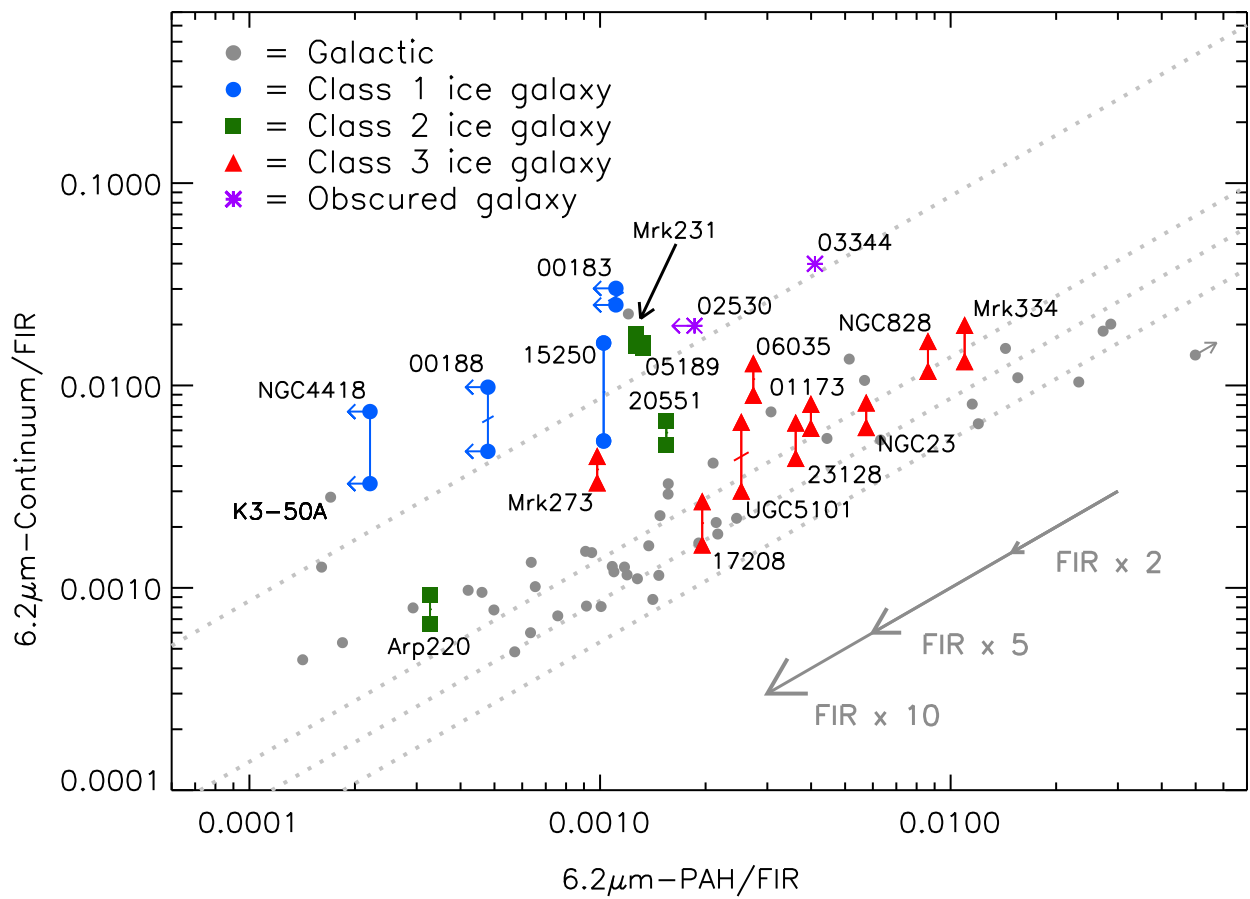}}}
\end{center}
\caption{MIR/FIR diagnostic diagram for deeply obscured galaxies and
ice galaxies.  The {\it dotted lines} and the {\it arrows} are as in
Fig.~\ref{contpahfir_galactic}. A vertical {\it line} indicates
the upward shift of an ice galaxy if the measured 6.2\,$\mu$m
continuum is corrected for the presence of 6.0\,$\mu$m water ice
absorption.}
\label{contpahfir_icegal}
\end{figure}

\clearpage

Fig.\,\ref{contpahfir_icegal} shows the distribution of a subset of
galaxies in which the 6.0\,$\mu$m water ice feature has been
identified \citep{Spoon:silenpahs:01}, the so-called `ice galaxies'.
The sample has been appended with I\,03344--2103 and
I\,02530+0211, which show similarly deep silicate absorption features
as the class 1 and class 2 ice galaxies. However, shortward of the
silicate feature both spectra show features which are likely
artifacts.  For I\,02530+0211 it is nevertheless clear that the
5.8--6.5\,$\mu$m spectrum is feature-free and for
I\,03344--2103 the presence of weak 6.2\,$\mu$m PAH emission can be
confirmed; the presence of a 6.0\,$\mu$m water ice absorption feature,
however, cannot be confirmed.

Generally, ice extinction is weak and has little influence on these
ratios. However, assuming that the PAH emission arises from a
circumnuclear region which is unaffected by extinction, strong ice
band sources would shift up by a factor of 2--3
\citep{Spoon:silenpahs:01}.

Class 3 ice galaxies, showing weak 6.0\,$\mu$m ice band absorptions,
coincide with the location of the normal and starburst galaxies in
this diagram.  Compared to class 3, class 1 ice galaxies, with no
obvious PAH features and strong 6.0\,$\mu$m ice and silicate bands are
shifted upwards and to the left in Fig.\,\ref{contpahfir_icegal},
betraying the addition of extra MIR continuum as well as FIR
continuum. In particular, we note that the extreme example of this
class, NGC\,4418, is located close to K3-50~A. The few class 2 ice
galaxies, with weak PAHs and moderate ice absorptions, are more
scattered throughout this plot. Some (I\,20551--4250) fall in between
the two other classes.  Others coincide with class 1 (Mrk\,231 and
I\,05189--2524). Arp\,220, which has weak ice absorption, has shifted
downwards along the Galactic trend and is an outlier in this diagram
(see above). It is clear that the absorption associated with molecular
clouds can influence the location of a region in this diagram.

In \citet{Spoon:silenpahs:01}, we proposed that the classification of
ice galaxy spectra may reflect an evolutionary sequence from strongly
obscured beginnings of star formation (and AGN activity) to a less
enshrouded stage of advanced star formation (and AGN activity).
Assuming NGC\,4418 to represent the earliest, most obscured stage
after a galaxy collision or merger, evolution would then
proceed towards the lower distribution for galaxies which in the MIR
are dominated by star formation (and may contain a Sf2 nucleus),
or towards the upper distribution for galaxies which in the MIR are
AGN-dominated.

\clearpage

\begin{figure*}[!t]
\begin{center}
\resizebox{\hsize}{!}{{\includegraphics{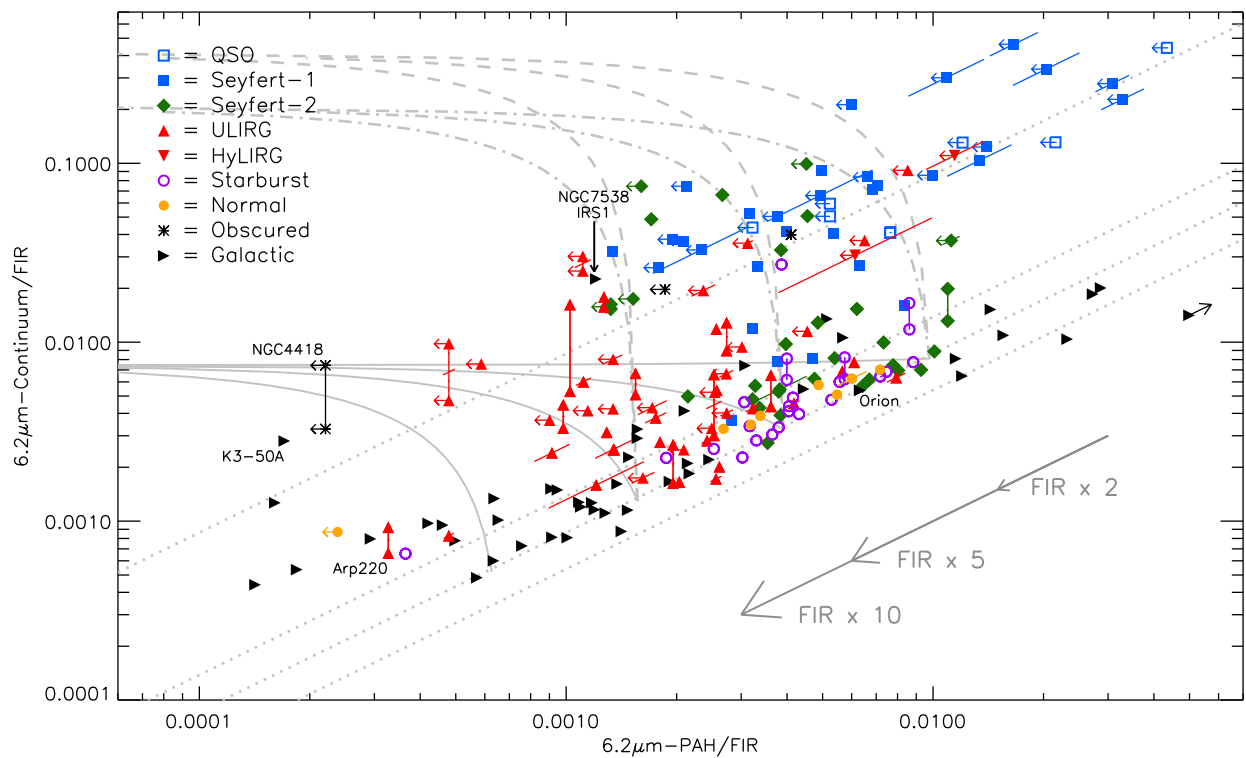}}}
\end{center}
\caption[]{The MIR/FIR diagnostic diagram for all sources. The {\it
dotted lines} and the {\it arrows} are as in
Fig.\,\ref{contpahfir_galactic}. The {\it continuous lines} are mixing
lines between heavily obscured and exposed star formation. The mixing
lines from starburst- to AGN-dominated MIR spectra are represented by
the {\it dashed} and {\it dot-dashed lines}, assuming respectively
25\% and 50\% of the FIR flux to be associated with AGN hot
dust. See text for details.}
\label{contpahfir_diagram}
\end{figure*}

\clearpage

\subsubsection{Application of diagnostic}
\label{application_plotje_zonder_naam}

MIR galaxy spectra can show the spectral characteristics of star
forming regions, AGN-heated dust and/or obscuration
(Sect.~\ref{spchar_galaxies}).  As pointed out above, spectra of clear
lines of sight to AGN-heated dust are located in other parts of the
MIR/FIR diagnostic diagram than spectra for which this line of sight
is blocked by the AGN torus and which are therefore PAH-dominated. In
order to better understand the connection between these two groups we
have calculated the diagnostic ratios that result from adding AGN hot
dust spectra to several choices of PAH-dominated spectra. For a pure
AGN-heated dust template, we used the nuclear spectrum of the
prototypical Sf1, NGC\,4151, and set its 6.2PAH/FIR ratio equal to
0. As the FIR flux associated with AGN hot dust has not been directly
measured for NGC\,4151 (nor for any other galaxy), we determine two
values for its 6.2cont/FIR ratio, assuming 25\% and 50\% of the FIR
flux to be associated with the AGN hot dust. For the starburst
template, we choose the central region of the starburst galaxy
M\,83. Other PAH dominated galaxy spectra along the fundamental line
are then mimicked by scaling its FIR flux up and down by a factor
3. Dashed and dash-dotted lines in Fig.~\ref{contpahfir_diagram}
represent the calculated mixing lines, further referred to as
`AGN-starburst' tracks.  The tracks strikingly illustrate the
direction in which a Seyfert galaxy spectrum moves if the orientation
of the AGN torus changed in order to have less obscuring dust in the
line of sight to the hot inner wall of the torus. Alternatively, they
illustrate the effect of an increase in AGN power. The absence of
Seyfert galaxies along the horizontal part of these tracks is due to a
combination of ISO's inability of detecting the 'weak' PAH emission on
such strong continua and the non-negligible contribution of star
formation from the host galaxy.

Similarly, starburst and heavily obscured galaxies are found in
far-apart regions in the MIR/FIR diagnostic diagram. We illustrate the
connection between these two groups by calculating the diagnostic
ratios that result from adding PAH dominated spectra to that of an
heavily obscured nucleus (NGC~4418). In this way, we mimic the degree
of obscuration.  We further used the same starburst templates as for
the AGN-starburst tracks, but added another fundamental line
`starting' point in the lower left quadrant, by scaling the FIR flux
of M\,83 up by a factor 30. These tracks -- further referred to as
`obscured-starburst' tracks -- are shown by solid lines in
Fig.~\ref{contpahfir_diagram}. Note that \HII regions located above
the fundamental line do not necessarily suffer from obscuration
(Sect.~\ref{mw_plotje_zonder_naam}).

A third set of mixing lines connects heavily obscured nuclei
(e.g. NGC\,4418) and galaxies dominated by AGN hot dust (e.g.
NGC\,4151). Assuming no PAH emission from either type of source (as
for the other mixing lines) implies these tracks would run vertical.
Taking into account a contribution of star formation to the spectra of
Seyferts will cause these tracks to run almost parallel to the lines
of constant 6.2PAH/6.2cont ratio. As several ULIRGs with signs of
strong obscuration (i.e. I\,00183--7111, Mrk\,231 and I\,00275--2859)
are lined up along these mixing lines, these three ULIRGs may well
harbour a hidden AGN.
 
No galaxies are found with 6.2PAH/FIR ratios below 0.002--0.003 on the
main sequence (Fig.~\ref{contpahfir_diagram}). Assuming this to be the
lowest attainable 6.2PAH/FIR value for pure starburst or normal
galaxies, our tracks suggest that galaxies found at 6.2PAH/FIR$<$0.002
likely contain an obscured nuclear component.  Interestingly, this
would then indicate that a large fraction of the ULIRG nuclei is
obscured. 

Note that our simple mixing model does not offer an
explanation for the extreme positions of Arp\,220, IC\,860,
CGCG\,049--057 and I\,03521+0028, found in the lower left quadrant of
the diagram.

\subsection{Laurent diagnostic diagram}
\label{laurent}

Another diagnostic solely based on the MIR spectrum is proposed by
\citet{Laurent00}. These authors assume that the MIR emission of
galaxies originates in 1) AGN-heated dust, characterized by a strong
dust continuum shortwards of 10 \mum, 2) PDRs, characterized by PAH
emission, and 3) \HII regions, characterized by strong dust continuum
longwards of 10 \mum. To quantify their relative contributions, two
diagnostic indicators are used, the ratio of warm (14--15 \mum) to hot
(5.1--6.8 \mum) dust continuum and the ratio of 6.2 \mum\, PAH
emission to hot dust continuum. Note that the latter indicator is
similar to the offset (6.2PAH/6.2cont) of a source from the
fundamental line in the MIR/FIR diagnostic.

\clearpage

\begin{figure*}[!t]
\begin{center}
\resizebox{\hsize}{!}{{\includegraphics{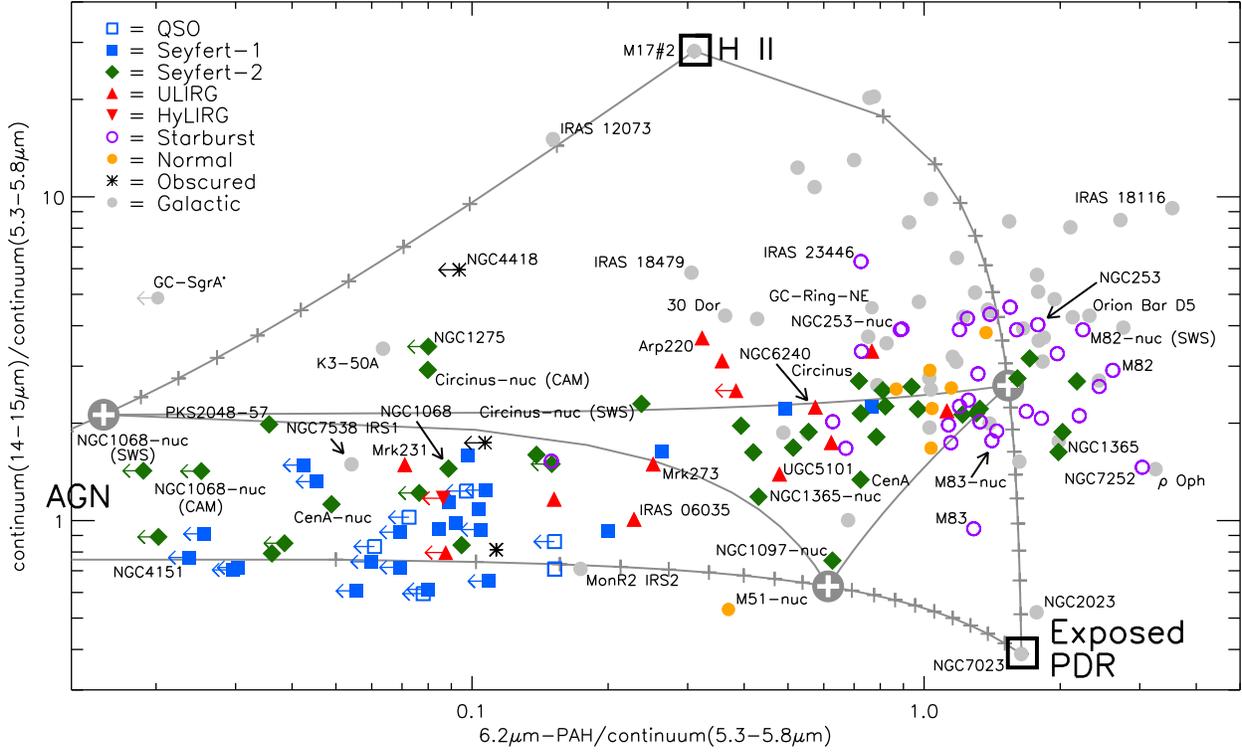}}}
\end{center}
\caption[]{MIR diagnostic diagram adapted from \citet{Laurent00}. The
three templates are indicated by their name (PDR, HII or AGN) and a
large open square. Since the AGN template has per definition no PAH
flux, it falls off the plotted x-range (at the left).  The {\it grey
lines} connecting the three templates are mixing lines
between the templates. Each {\it grey cross} indicates a 5\% change in
relative contribution. {\it Thick crosses} mark equal contribution of
two templates. The {\it grey lines} connecting the three {\it thick
crosses} indicate a constant 50\% fraction of one component along each
line. The SMC source B1\#1 is located off the diagram at position
(3.0,0.16).}
\label{laurent_diagram}
\end{figure*}

\clearpage

We choose to sample the hot continuum at shorter wavelengths
(i.e. 5.3--5.8\,$\mu$m; referred to as 5.5cont) than \cite{Laurent00},
in order to be a) more sensitive to the slope between hot and warm
continuum and b) to avoid sampling within the 6.0\,$\mu$m ice
absorption band. For the warm continuum we adopt the same integration
range as \cite{Laurent00}: 14.0--15.0\,$\mu$m, referred to as 15cont.
The 6.2 \mum\, PAH flux is determined as defined in
Sect.~\ref{plotje_zonder_naam}.  The derived fluxes for the Galactic
sample and the extragalactic sample are given in, respectively,
Table~\ref{log} and Spoon et al. (in prep.).

The selection criteria for this diagram are somewhat different than
for the previous diagram (Sect.~\ref{plotje_zonder_naam}), as not all
galaxy spectra are available over the full 5.3--15.0\,$\mu$m
range. ISO--PHT--S spectra, for instance, only cover the ranges
2.47--4.77\,$\mu$m and 5.84--11.62\,$\mu$m. For targets at low
redshift this means that the 5.5cont cannot be measured directly. We
therefore performed a linear interpolation over the 4.77--5.84\,$\mu$m
gap for those spectra which have sufficient S/N on both sides. Targets
which did not meet this criterion, mostly ULIRGs, were eliminated from
the sample. For sources with redshifts $z$$>$0.1 the gap in the
ISO--PHT--S spectral coverage is not an issue. ISO--PHT--S spectra
also present a problem at the long wavelength end, as their wavelength
coverage ends at 11.62\,$\mu$m. In order to derive a 15cont for these
cases, we logarithmically interpolated the IRAS 12\,$\mu$m and
25\,$\mu$m fluxes to the 14--15\,$\mu$m range.  However, sources were
removed from the sample if their IRAS 12\,$\mu$m fluxes are upper
limits, or if there is a clear mismatch between the ISO--PHT--S
spectrum and the IRAS 12\,$\mu$m flux. While the former criterion
results in the exclusion of many fainter ULIRG spectra, the latter
criterion results in the non-selection of many nearby galaxies.  We then
screened against sources with 5.3--5.8\,$\mu$m spectra which are
either too noisy or which show artifacts, as is the case for several
ISO--CAM--CVF spectra.  Our final galaxy sample consists of 66 AGNs (6
QSOs, 23 Sf1s and 37 Sf2s), 30 starburst and \HII-type galaxies, 7
normal galaxies, 13 ULIRGs, one HyLIRG and 3 IR galaxies with deeply
obscured nuclei. Note that some galaxies are close enough for ISO to
obtain spectra of both the central region and the entire ISO--CAM--CVF
field of view. These galaxies are NGC\,253, NGC\,1068, NGC\,1365,
NGC\,1808, Cen\,A, Circinus and M\,83.

\subsubsection{\HII regions and ISM}

The \HII regions do not form a narrow sequence in this diagram but
show a wide spread scattering over about 1/3 of the plot. However,
they do avoid the low PAH and low warm continuum properties of the
AGNs. Correcting for extinction (for sources with known A$_K$) does
not reduce the observed spread. However, it clearly has some influence
on the precise position of the sources; an A$_K$ of 2, which is
typical for these \HII regions \citep{paperii, Martin:radio:02}, gives
rise to an increase of 7\% and 15\% in 6.2PAH/5.5cont and
15cont/5.5cont, respectively. The Orion Bar is now positioned with the
\HII regions instead of with the RNe
(Sect.~\ref{mw_plotje_zonder_naam}). Obviously, Galactic sources
offset from the fundamental line in the MIR/FIR diagram (i.e. with
unusual 6.2PAH/6.2cont ratios) are also outliers in this diagram
(i.e. at unusual 6.2PAH/5.5cont ratios) but are easier notable here.
Hence, similar conclusion can be drawn as for the MIR/FIR
diagnostic. Highly embedded regions (NGC7538\,IRS1, Mon\,R2\,IRS2, and
K3--50A) are found at low 6.2PAH/5.5cont and low 15cont/5.5cont,
clearly away from \HII regions and much closer to the AGN template.

\subsubsection{Galaxies}

The majority of galaxies are found in a wide strip, almost
horizontally (within a factor of 3) across the diagram
(Fig.\,\ref{laurent_diagram}). This sequence runs from normal and
starburst galaxies on the right (coinciding with the \HII regions),
through the ULIRGs in the middle to the Sf1s on the left. Sf2s are
predominantly located on the right, but are found along the full
range, depending on the degree of dilution of the AGN hot dust
continuum by a PAH dominated spectrum. This depends on both the
physical size covered by the aperture and the fraction of the
bolometric luminosity contributed by star formation.  We note the
close similarity between the MIR characteristics of the Orion Bar and
the nucleus of starburst galaxy M\,82. In contrast, the spectrum of the entire
M\,82 field-of-view is closer to that of RNe. The same
trend is observed for M\,83. Apparently, galaxy disks are dominated by
stellar types slightly more similar to RNe.
Fig.\,\ref{laurent_diagram} reveals small differences in continuum
slope among AGN nuclei, for example between the nuclear spectra of
NGC\,1068 (Sf2) and NGC\,4151 (Sf1). As pointed out by
\cite{Laurent00}, the spectra of the AGN host galaxies are
dramatically different from those of the AGN itself and resemble those
of normal and starburst galaxies instead. A good example is the nearby
Sf2 Cen\,A \citep{Laurent00}. While its integrated galaxy spectrum is
PAH dominated and is found among the starburst galaxies, the nuclear
spectrum shows hardly any PAH emission and is situated with the
AGNs. Unlike the MIR/FIR diagram, where ULIRGs occupy their own niche,
ULIRGs are not separated out as much. Some ULIRGs (Mrk\,231 and
I\,23060+0505) and the HyLIRG I\,09104+4109 are found among the pure
AGNs, while other ULIRGs, like I\,17208--0014, are found among the
starburst galaxies. The remaining ULIRGs are found more towards the
center of the plot. The highly obscured source NGC\,4418 has a
ratio of 15cont/5.5cont which is very similar to that of the
prototypical starburst galaxy M\,82. However, it lacks the strong 6.2
\mum\, PAH feature of M\,82 and hence its 6.2PAH/5.5cont ratio is typical
of pure AGN spectra; the slope of its continuum is however much steeper.
Other sources with moderate to strong obscuration are found in
the lower left quadrant, overlapping to some extent with the Galactic
embedded star forming regions (Fig\,\ref{laurent_diagram}).  Note that
Arp\,220 is found in between NGC\,4418 and the average position of
starburst galaxies, in agreement with the results of
\citet{Spoon:arp220:04}.

\subsubsection{Application of diagnostic}

Similar to \citet{Laurent00}, we calculated mixing lines where the
contribution of each template to the MIR flux (measured from 5.3 to 15
\mum) varies between 0 and 100\% (Fig.~\ref{laurent_diagram}).  The
three templates used, are the spectra of NGC\,7023 (exposed PDR),
M\,17 at position 2 (\HII region) and NGC\,4151 (AGN-heated dust). The
templates are represented in Fig.~\ref{laurent_diagram} by large open
squares except for the AGN template for which we assumed a PAH flux of
zero instead of the derived upper limit.  It is important to realize
that the mixing percentages can only be interpreted as percentages of
{\it MIR} and not {\it total} IR luminosity. This is especially true
for the AGN contribution, as the same AGN may have a factor 10 higher
or lower continuum flux, solely depending on the degree of obscuration
towards the central source.

\clearpage

\begin{figure*}[!t]
\begin{center}
{\includegraphics[height=7cm]{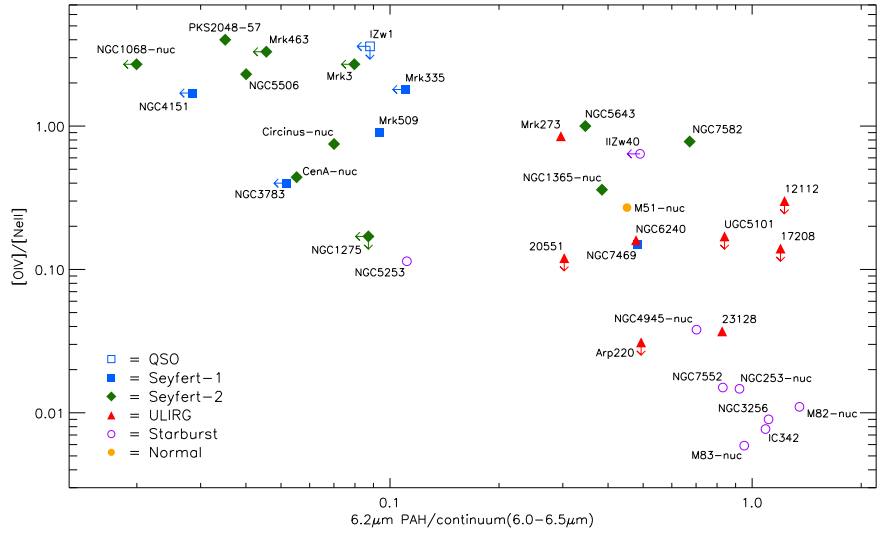}}
\end{center}
\caption[]{MIR diagnostic diagram for different type of galaxies,
adapted from \cite{Genzel98}. AGN-dominated MIR spectra are found in
the upper left quadrant, while starburst-dominated MIR spectra are
found in the lower right quadrant.}
\label{genzel_diagram}
\end{figure*}

\clearpage

Our sample of \HII regions is clearly not just dominated by
emission of hot dust from the \HII region but also shows significant
contribution from the surrounding PDR. The sequence from PDRs
to \HII regions is not well determined observationally. Using other
templates to calculate the mixing lines would change its position.
Indeed, this is forcefully brought home by the many objects which lie
outside the mixing lines which span the diagram. Changing the
templates will not remove the spread of the observed \HII regions
along a mixing line. Hence, \HII regions can only be described in
first order by a combination of an \HII region and PDR template.

Embedded protostars are easily distinguished from \HII regions and
PDRs in this diagram by their lower 6.2PAH/5.5cont
ratio. Unfortunately, their location coincides with that of
AGN-dominated spectra. Hence, this diagnostic diagram cannot
distinguish between MIR spectra dominated by an AGN or
by a deeply embedded source.  This is of particular interest
for ULIRGs whose MIR spectra have the signature of star formation,
AGN-heated dust or dust extinction (Sect.~\ref{spchar_galaxies}). With
this diagram, only the first can be discerned from the rest.

The PAH contribution to the MIR spectrum of Seyferts is best traced by
the 6.2PAH/5.5cont ratio, which is highly sensitive to small fractions
of (exposed) PDR contribution included within the beam. These
contributions usually arise from circumnuclear star formation rings or
from the galaxy disk.  Note that for starburst galaxies the inclusion
of a larger part of the galaxy disk does not result in a similar
strong shift as for Seyferts in either diagnostic ratio.

\subsection{The Genzel diagnostic diagram}
\label{genzel}

A MIR diagnostic diagram which also successfully separates Seyfert
galaxies from starburst galaxies is the diagram proposed by
\cite{Genzel98}. As seen in Sects.~\ref{plotje_zonder_naam}
and~\ref{laurent}, the use of the 6.2PAH/6.2cont (or equivalently the
7.7 \mum\, PAH line-to-continuum) is an excellent choice for
determining the importance of star formation in the MIR. Likewise, the
ratio of the high excitation fine-structure line
25.9\,$\mu$m\,[\ion{O}{4}] to the low excitation fine-structure line
12.8\,$\mu$m\,[\ion{Ne}{2}] is a good tracer for AGN activity.  Here
we show a modified version of this diagram, in which we replaced the
7.7 \mum\, PAH line-to-continuum ratio by the 6.2PAH/6.2cont
(Fig.\,\ref{genzel_diagram}). We note that our Galactic sample does
not exhibit the high excitation fine-structure line
25.9\,$\mu$m\,[\ion{O}{4}] emission.

The number of galaxies in our Genzel diagram is limited by the
availability of line fluxes of 12.81\,$\mu$m [\ion{Ne}{2}] and
25.9\,$\mu$m [\ion{O}{4}] from the literature. For AGNs the line
fluxes were taken from \cite{Sturm02}, for starburst galaxies from
\cite{Verma03} and for ULIRGs from \cite{Genzel98}. Our final sample
consists of 17 AGNs (1 QSO, 5 Sf1s, 11 Sf2s), 8 ULIRGs, 9
starburst galaxies and 1 normal galaxy.

The [\ion{O}{4}]/[\ion{Ne}{2}] ratio by itself is effective in
separating Seyfert (both Sf1s and Sf2s) from starburst galaxies, as
Seyferts have ratios higher than 0.1 and starburst galaxies have
ratios lower than $\sim$0.02 (Fig.\,\ref{genzel_diagram}). Exceptions
are low metallicity galaxies, e.g. NGC~5253 and IIZw40.  The other
diagnostic ratio, 6.2PAH/6.2cont, is not a reliable separator of Sf1s
and Sf2s on one hand and starburst galaxies on the other hand. This
was already discussed in the context of the MIR/FIR diagram and is
strikingly illustrated in the Genzel diagram by the examples of the
Seyferts NGC\,7469 and NGC\,7582, which appear starburst-like in their
6.2PAH/6.2cont ratios, but which are AGN-like in their [OIV]/[NeII]
ratios.  Other Seyferts would show a similar disagreement in their
diagnostic ratios, had their properties been represented by their
integrated galaxy spectra instead of their nuclear spectra. Examples
are Cen\,A and Circinus. Based on the above, we conclude that the
[OIV]/[NeII] ratio is a more reliable AGN indicator than the
6.2PAH/6.2cont.  Applying the [\ion{O}{4}]/[\ion{Ne}{2}] ratio to find
AGNs in ULIRGs, only 2 out of 8 ULIRGs harbor an AGN (NGC6240 and
Mrk\,273). The other 6 ULIRGs either have upper limits or have a ratio
intermediate to those typical for Seyferts and starbursts
(I\,23128--5919).

The nucleus of the nearby starburst galaxy NGC\,4945 is found at an
intermediate [\ion{O}{4}]/[\ion{Ne}{2}] ratio. The elevated
[\ion{O}{4}]/[\ion{Ne}{2}] ratio for this source is likely due to
strong differential extinction between 12.81\,$\mu$m and 25.9\,$\mu$m
in this notoriously dusty nucleus \citep{Spoon:ngc4945:00}.  Hard
X-ray observations have shown this galaxy to contain a buried AGN
\citep{Iwasawa93,Guainazzi00}, which so far has escaped detection at
NIR and MIR wavelengths.  Therefore, NGC\,4945 may be taken as a
warning that some AGNs may escape detection also from MIR excitation
indicators like [\ion{O}{4}]/[\ion{Ne}{2}].  Other highly obscured
galactic nuclei, like NGC\,4418 and Mrk\,231, have not even been
detected in 12.81\,$\mu$m [\ion{Ne}{2}] \citep{Spoon:ngc4418:01,
Genzel98}. For these galaxies, and likely also for many ULIRGs, the
Genzel diagnostic diagram is not well suited.

\subsection{Comparison of the three diagnostic diagrams}

Each of the three diagnostic diagrams is constructed with the
immediate goal to reveal the identity of a galaxy.

All three diagnostics are able to identify Sf1s. The degeneracy
between starburst and Sf2s can only be broken by the Genzel diagram.
Indeed, by using the [\ion{O}{4}]/[\ion{Ne}{2}] ratio, the Seyferts
are clearly separated from the starburst galaxies. In contrast,
starburst galaxies and Sf2s occupy the same region in the MIR/FIR and
Laurent diagnostic diagram. This illustrates that the identification
of the dominant power source --- AGN or starburst --- highly depends
on the wavelength region considered. Conversely, such an
identification does not imply a similar degree of dominance in the
{\it total} IR luminosity.  

Heavily obscured galaxies are best recognized in the MIR/FIR
diagram. Since they show no or lack detectable fine-structure lines,
they are absent in the Genzel diagram, while in the Laurent diagram
--- which was also not constructed for these type of galaxies --- they
can be mistaken for AGN-dominated sources. Only the MIR/FIR diagram
seems to be able to separate out ULIRGs showing signatures of obscured
star formation or obscured AGN activity.

\section{Discussion}
\label{discussion}

\subsection{PAH abundance}
\label{abundance}

The ratio of the total PAH emission to the FIR emission measures the
competition of the PAHs and the dust for the UV photons and is, thus,
an indicator for the PAH/dust abundance.  Studies of PAH/FIR ratios in
Galactic sources have revealed that this ratio is independent of the
local radiation field, G$_0$, for low radiation fields and decreases
with G$_0$ at higher radiation fields \citep{Boulanger:98,
Boulanger:leshouches:98, Onaka:00}. These authors suggested that this
proportionality of PAH strength with G$_0$ for low G$_0$ (since
PAH/FIR is constant with G$_0$) is consistent with emission coming
from species small enough to be stochastically heated. The decrease
for high G$_0$ is then taken to indicate a decreasing abundance of
these species relative to that of the grains with increasing strength
of the illuminating radiation field.

\clearpage

\begin{figure}[t!]
\begin{center}
\resizebox{\hsize}{!}{{\includegraphics{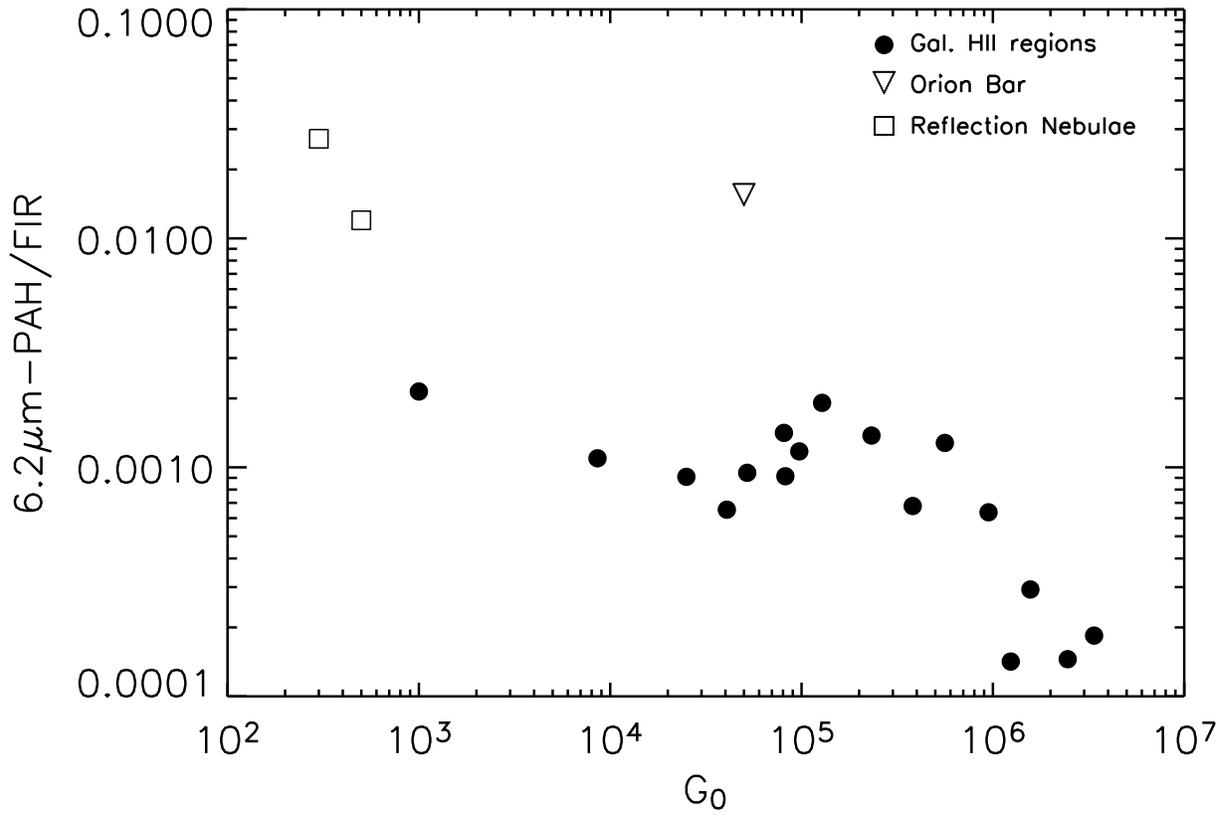}}}
\end{center}
\caption{The relation of 6.2PAH/FIR with the local radiation field,
G$_0$ for \HII regions. As a reference, two RNe are also shown. }
\label{g0}
\end{figure}


\clearpage

To first order, the PAH/FIR can be traced by the 6.2PAH/FIR since the
fraction of total PAH flux emitted in the 6.2 \mum\, PAH band varies
only from 14 to 38 \% with an average of 28$\pm$4\%
\citep{Vermeij:pahs:01, Peeters02b}.  We derived $G_0$ values from the
IR flux and the angular size of the PAH emission region
\citep[cf. ][]{Hony:oops:01}. This assumes that all the UV light is
absorbed in a spherical shell with the angular diameter of the \HII
region and re-emitted in the IR. Since the PAHs are expected to be
destroyed inside the \HII region, it is reasonable to use the radio
size of the \HII regions (taken from \citet{Peeters:cataloog:01} and
\citet{Martin:radio:02}). For $G_0$ values for the Orion bar and
the RNe, we refer to \cite{Tielens:anatomyorionbar:93},
\citet{Joblin:3umvsmethyl:96}, and \citet{Hony:oops:01}. Extended and
complex \HII regions are excluded due to possible aperture effects.

Fig.~\ref{g0} shows the relation between 6.2PAH/FIR and G$_0$ for our
sample of \HII regions. Taking into account the results of
\citet{Boulanger:98, Boulanger:leshouches:98} for G$_0 <$ 10$^3$, this
relation seems to be a step function with steps at G$_0$ $\sim$10$^3$
and $\sim$10$^6$, exposing the Orion Bar as a clear outlier.  Indeed,
sources with similar 6.2PAH/FIR show a large spread in $G_0$, up to a
factor 10$^3$. This first step was seen by \citet{Boulanger:98,
Boulanger:leshouches:98} based on the difference between regions with
low G$_0$ and the \HII region M17. In contrast, \citet{Onaka:00} found
a loose correlation between 7.7PAH/FIR versus G$_0$ for different
positions within the Carina Nebula (with only a factor up to 7
difference in G$_0$ for positions with similar 6.2PAH/FIR).  The
presence of such a relationship within a single object seems to be a
more general characteristic of the PAH emission behavior. Apparently,
within a single object, variations in G$_0$ are important in driving
the emission spectrum. However, the source to source variation in the
PAH/FIR ratio does not seem to follow G$_0$, but is rather dominated
by other factors.  Fig.~\ref{g0} also demonstrates that PAHs compete
much better for FUV photons in the diffuse ISM than in compact \HII
regions. This may reflect the destruction of PAHs inside the ionized
gas volume since dust can be present inside \HII regions
\citep{Martin:radio:02} and thereby absorb much of the FUV flux before
it even reaches the surrounding PAH-rich PDRs. In this case, the
importance of this PAH destruction in the ionized gas will increase
with decreasing size of the \HII region because this corresponds to a
larger dust optical depth in the ionized gas for a given dust
abundance and hence absorb more of the FUV flux.  Possibly, the
observed large spread in the 6.2PAH/FIR ratio with G$_0$ reflects the
variation in the internal dust content of \HII regions. It is fair to
say, however, that at this point the origin of the large variation in
the 6.2PAH/FIR ratio is unclear.

\subsection{PAHs as a tracer of star formation}

Star formation properties of galaxies are essential in assessing their
evolutionary histories. Different tracers for star formation are used
based upon integrated colors, the UV continuum, recombination and
forbidden lines and FIR emission \citep[e.g.][and references
therein]{Kennicutt:sfr:98}. PAHs may also provide a convenient tracer
of star formation activity as Sect.~\ref{diagnostictools}
exemplifies. PAHs are stochastically heated mainly by UV photons
produced by massive stars \citep[PAHs can also be excited by visual
photons, though the excitation is dominated by UV
photons, e.g.][]{Uchida:98, Li:02}. Assuming fixed emission and
absorption properties and fixed PAH abundance, the PAH emission is a
measure of the amount of photons available between 6 and 13.6 eV (the
former corresponding with the averaged ionization potential of PAHs)
and hence of star formation.  However, using the PAH emission as a
tracer of star formation activity is wrought with difficulty as
Sect.~\ref{abundance} illustrates; i.e. the 6.2PAH/FIR varies over 2
orders of magnitude for Galactic \HII regions in our sample.  To
assess the robustness of this tracer, we compared it to other star
formation tracers, N$_{\rm lyc}$ and L$_{\rm FIR}$, both for Galactic
star forming regions and galaxies (normal and starbursts).

\clearpage

\begin{figure}[t!]
\epsscale{0.7}
\plotone{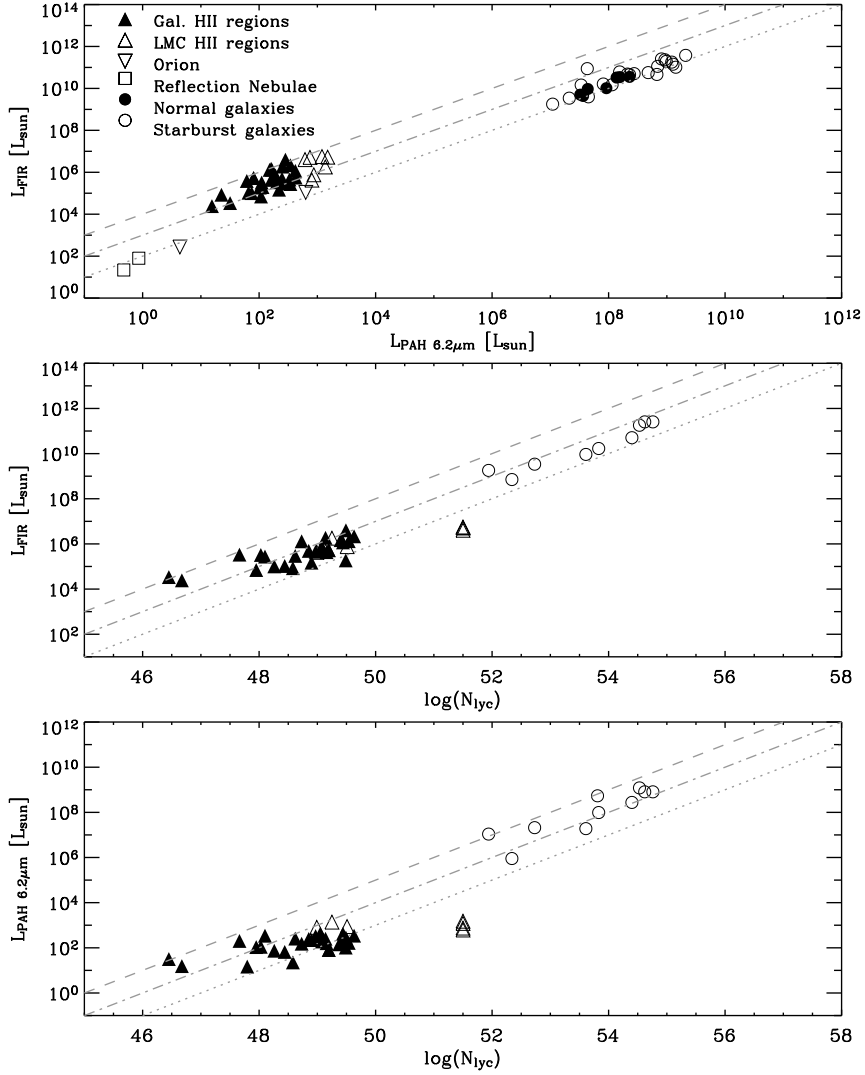}
\caption{Three tracers of star formation, L$_{6.2 PAH}$, log(N$_{\rm
lyc}$) and L$_{FIR}$ plotted against each other for Galactic \HII
regions, LMC \HII regions \citep{Vermeij:pahs:01} and normal and
starburst galaxies. In case of a distance ambiguity for the Galactic
\HII regions, only the far distance is shown. The {\it grey lines}
({\it dotted, dash-dotted, dashed}) indicate a L$_{6.2 PAH}$ equal to
respectively 1, 0.1 and 0.01 \% of L$_{FIR}$ ({\bf top panel}), a
N$_{\rm lyc}$/L$_{\rm FIR}$ ratio of respectively 10$^{44}$, 10$^{43}$
and 10$^{42}$ ({\bf middle panel}) and a N$_{\rm lyc}$/L$_{\rm
6.2PAH}$ ratio of respectively 10$^{47}$, 10$^{46}$ and 10$^{45}$
({\bf bottom panel}).  }
\label{nlyc}
\end{figure}

\clearpage

L$_{\rm FIR}$ traces star formation since a significant fraction of
the stellar radiation is emitted by (young) stars with spectral types
mid-B or earlier and is absorbed by dust and re-emitted thermally in
the FIR.  Also, the number of ionizing photons, N$_{\rm lyc}$, is
commonly used as a measure of massive star formation. It is mainly
derived in three ways, 1) from the H$\alpha$ recombination line; 2)
from radio recombination lines and 3) from the hydrogen free-free
continuum emission at radio wavelengths. For our Galactic and LMC \HII
regions, N$_{\rm lyc}$ is obtained from radio continuum emission
\citep{paperii, Martin:radio:02, Vermeij:pahs:01}. For our galaxies,
free-free radio emission may be contaminated with non-thermal
synchrotron emission from young SNR or AGN activity. Their N$_{\rm lyc}$ is
therefore obtained from near- and mid-IR hydrogen recombination lines
instead \citep{Genzel98, Verma03, Foerster01}.

L$_{\rm FIR}$ is in first order (on this scale -- within a factor of
100) clearly linearly proportional to L$_{\rm 6.2PAH}$
(Fig.~\ref{nlyc}).  However, their ratio is significant different
between the Galactic \HII regions (5.5$\pm$4.6~10$^{-4}$) and the
galaxies (6.3$\pm$3.2~10$^{-3}$), with Orion and the RNe having a
similar and lower ratio as the galaxies, respectively (also seen in
Figs.~\ref{contpahfir_galactic} through~\ref{contpahfir_diagram}).  In
contrast, this discrepancy between the Galactic and extragalactic
sample is not present in the L$_{\rm FIR}$/N$_{\rm lyc}$ ratio, though
a slightly larger scatter is observed in each of these groups
(Fig.~\ref{nlyc}). In addition, L$_{\rm 6.2PAH}$ is roughly
proportional to N$_{\rm lyc}$ for the galaxies \citep[consistent
with][]{Roussel:01} while for the Galactic \HII regions, it is much
less dependent on N$_{\rm lyc}$ (Fig.~\ref{nlyc}). In fact, the
L$_{\rm 6.2PAH}$ observed in \HII regions seems to be almost
independent of the hardness of the radiation field.

\begin{figure}[t!]
\begin{center}
\resizebox{\hsize}{!}{{\includegraphics{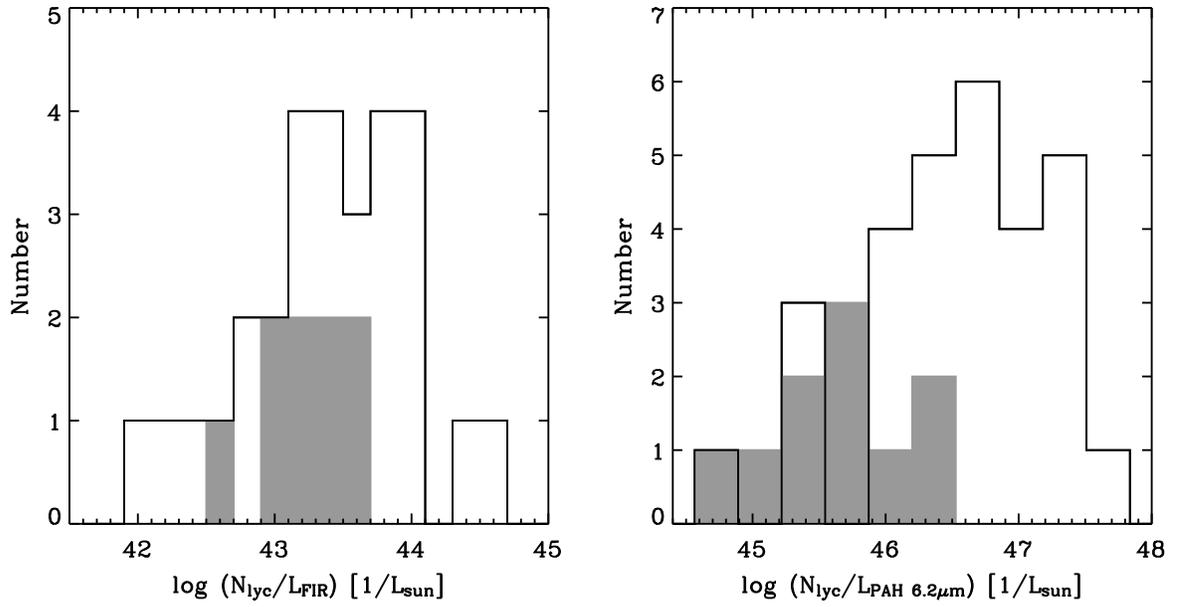}}}
\end{center}
\caption{The distributions in N$_{\rm lyc}$/L$_{\rm FIR}$ and N$_{\rm
lyc}$/L$_{\rm 6.2PAH}$ for the Galactic and LMC \HII regions ({\it
solid line}) and the normal and starburst galaxies ({\it grey
scale}). }
\label{nlyc_histo}
\end{figure}

As discussed by e.g. \citet{Kennicutt:sfr:98}, and references therein,
the robustness of L$_{\rm FIR}$ as a tracer for star formation depends
on the type of galaxies considered, being highest for dusty
circumnuclear starbursts. Indeed, the FIR spectra of galaxies are
composed of emission of dust around young star-forming regions and
emission of more extended dust heated by the interstellar radiation
field.  The same holds for PAHs as a tracer of star
formation. Likewise, the observed PAH flux is integrated over the
whole galaxy and therefore also includes PAH emission originating in
the ISM, RNe, exposed PDRs and embedded compact \HII regions. This ISM
contribution can be estimated by comparing L$_{\rm 6.2PAH}$ with
N$_{\rm lyc}$ (since it does not suffer from this contamination) for
both the \HII regions --- as a template for massive star formation ---
and the galaxies. Concerning L$_{\rm FIR}$, the galaxies show a
similar distribution of N$_{\rm lyc}$/L$_{\rm FIR}$ as the \HII
regions (Fig.~\ref{nlyc_histo}) and so L$_{\rm FIR}$ is likely not
influenced by an ISM contribution in these galaxies. In contrast, the
galaxies have a clearly different distribution in N$_{\rm
lyc}$/L$_{\rm 6.2PAH}$ compared to the \HII regions, with --- on
average --- a lower ratio, indicating that the PAH emission in
galaxies partly originates in the ISM. Therefore, PAHs do not trace
massive star formation \citep[O stars; consistent with][]{Haas:02} and
may be better suited as a tracer of B stars, which dominate the
Galactic stellar energy budget. Obviously, PAHs are a bad tracer for
highly embedded massive star formation due to their absence in the MIR
spectrum of these objects.

\clearpage

\clearpage

\subsection{Conversion from PAH luminosity to IR luminosity}

Recently, \cite{Soifer02} and \cite{Lutz03} have used PAH emission as
a quantitative measure for the contribution of exposed star formation
to the bolometric luminosity of two ULIRGs (Arp\,220 and NGC\,6240).
While \cite{Soifer02} assumed the ratio of L(11.2PAH) and L(IR) for
the starburst core in M\,82 to be a measure for exposed star
formation, \cite{Lutz03} instead used the mean L(7.7PAH)/L(IR) ratio
derived from a sample of 10 starburst nuclei. A third method to derive
the bolometric correction was proposed in \citet{Spoon:arp220:04} and
is based on the mean L(6.2PAH)/L(IR) ratio for our sample of normal
and starburst nuclei.  This ratio is 3.4$\pm$1.7$\times 10^{-3}$.

For sources where the environment of massive star formation resembles
exposed PDRs \citep[such as M\,82 and NGC\,253;][]{Carral94,Lord96},
the three methods discussed above will give reasonable results. On the
other hand, for sources which resemble embedded star formation
(e.g. Arp\,220 and NGC\,4418), a ten times lower ratio, appropriate
for compact \HII regions like W3, might be a better choice.  Which
conversion ratio is more appropriate cannot be decided based upon the
mid-IR spectra alone.  Observations of PDR lines such as the
[\ion{O}{1}]\,63\,$\mu$m and [\ion{C}{2}]\,158\,$\mu$m lines can be
very instrumental in resolving this issue. In that respect, we note
that the PDR fine structure lines are very weak in Arp\,220
 and other ULIRGs \citep{Luhman03, Dale:04}, suggesting also that the PDR component is
underdeveloped, perhaps due to absorption of a major fraction of the
FUV flux by dust inside the \HII region \citep{Luhman03}.

\section{Conclusions}
\label{conclusions}

In this paper, the MIR spectral characteristics of Galactic
and extragalactic sources are investigated. Our sample includes
Galactic (C)\HII regions, ISM lines of sight and embedded massive
protostars as well as normal galaxies, starburst galaxies, Seyferts,
QSOs and (ultra-luminous) IR galaxies.  

First, the MIR spectrum of each object type is described and compared,
revealing distinct spectral characteristics for each object type.
In order to distinguish the different natures of the galaxies,
i.e. AGN-dominated, starburst-dominated or heavily obscured, we
present a new MIR/FIR diagnostic based on the ratio of the 6.2 PAH
emission band to FIR flux and the ratio of the 6.2\mum\, continuum to
FIR flux. This diagnostic is also applied to our Galactic sample. Both
ratios vary clearly within our sample of \HII regions and this range
extends up to the RNe and the (diffuse) ISM lines of sight. In
addition, the 6.2PAH/6.2cont ratio is varies slightly over a wide
range of Galactic \HII regions, as well as more general ISM
material. However, the observed variation in Galactic sources is much
smaller then that in galaxies. As such, it provides a very clear
handle on any AGN contribution to the MIR.  Indeed, AGNs are found to
segregate in two groups; most Sf2s are located with the normal and
starburst galaxies, while most Sf1s show strong 6.2cont/FIR
ratios. The 6.2 \mum\, PAH luminosity distributions are found to be
independent of the Seyfert type, in accordance with the orientation
dependent AGN unification scheme, and hence confirm the results of
\cite{Clavel00}.  This diagram further reveals the spectral
resemblance of starburst and normal galaxies to exposed PDRs rather
than (slightly embedded) compact \HII regions. ULIRGs show a diverse
spectral appearance. Some show a typical AGN hot dust continuum. More,
however, are either starburst-like or show signs of strong dust
obscuration in the nucleus.  One characteristic of the ULIRGs seems
also to be the presence of more prominent FIR emission than either
starburst galaxies or AGNs.
Comparison with the diagnostic diagrams proposed by \citet{Genzel98}
and \cite{Laurent00} for our sample (both Galactic and extragalactic)
indicates that the ability to identify obscured objects is best
achieved with our MIR/FIR diagnostic, while the ability to identify
the presence of an optically-recognized AGN is best achieved with the
Genzel diagram.

We found that the observed variation of the MIR/FIR diagnostic ratios
in the Galactic sample is linked with their evolutionary state and the
PAH/dust abundance. Finally, we have examined the use of PAHs as 
quantitative tracers of star formation activity and find that PAHs may 
be better suited as a tracer of B stars, which dominate the Galactic 
stellar energy budget, than as a tracer of massive star formation. 
Likewise, the PAH emission of normal and starburst galaxies is best 
represented by that of exposed PDRs such as the Orion Nebula.
However, the IR spectra of some sources --- notably the archetypal 
ULIRG Arp\,220 --- may be dominated by embedded massive star formation 
rather than exposed PDRs.

\acknowledgments

We thank D. Lutz and the anonymous referee for useful comments and
W.\,Reach, S.\,Madden and J.P.\,Simpson for providing spectra.  This
research is based on observations with ISO, an ESA project with
instruments funded by ESA Member States (especially the PI countries:
France, Germany, the Netherlands and the United Kingdom) and with the
participation of ISAS and NASA.  This research has made use of the
NASA/IPAC Extragalactic Database (NED), operated by the Jet Propulsion
Laboratory, California Institute of Technology, under contract with
the National Aeronautics and Space Administration.

\end{document}